\documentclass[prb,floatfix,10pt]{revtex4}
\usepackage{amsmath}
\usepackage[english]{babel}
\usepackage[pdftex]{graphicx}
\usepackage{color}

\usepackage{ulem}

\usepackage[sort&compress]{natbib}
\usepackage{hyperref}
\graphicspath{{images/}{./}}

\usepackage{amsmath}
\usepackage{amssymb}
\usepackage{mathtools}
\usepackage{braket}

\renewcommand{\vec}[1]{\ensuremath{\boldsymbol{#1}}}

\newcommand{\sgn}{\,\mbox{\rm sgn}}
\newcommand{\diag}{\,\mbox{\rm diag}}
\usepackage{amsmath}
\usepackage{afterpage}
\usepackage[english]{babel}
\usepackage[pdftex]{graphicx}
\usepackage{color}

\usepackage[sort&compress]{natbib}
\usepackage{hyperref}
\graphicspath{{images/}{./}}

\usepackage{amsmath}
\usepackage{amssymb}
\usepackage{mathtools}
\usepackage{braket}
\usepackage{float}

\usepackage{subfigure}

\usepackage{lipsum}
\begin{document}
\title{Pseudospin-one particles in the  time-periodic dice lattice: A new approach to transport control }

\author{Parisa Majari}
\email{majari@icf.unam.mx}
\affiliation{Instituto de Ciencias F\'isicas, Universidad Nacional Aut\'onoma de M\'exico, Cuernavaca 62210, M\'exico}

\begin{abstract}
The controlling of the transmission in the pseudospin-one Dirac-Weyl systems offers a rich tool to study new  concepts of massive Dirac electron tunneling by means of   a time-dependent potential. The  time-periodic potential is one of the experimental techniques to have more control over the tunneling effect. In this paper, we study the transmission coefficient for different sidebands to obtain total transmission. We show how the super  Klein tunneling  under special condition is  independent of  the incidence angle, oscillation amplitude, frequency, and barrier width. We  consider a band gap opening   with different locations of the flat band and  modulate the resonances  by tuning free parameters in our system.
\end{abstract}

\maketitle
\section{Introduction}
There has been a growing interest in two-dimensional (2D) materials since the first  synthesis of graphene in 2004 \cite{novoselov2004electric,de2007magnetic, ma2020investigation,stegmann2017transport,betancur2018controlling,sadurni2010playing,dell2018klein,moldovan2017magnetic,pena2015magnetostrain,hou2020valley}. Graphene is a one-atom thick crystal of hexagonal arrangement of carbon atoms which has remarkable properties such as high electronic and thermal conductivities, quantum Hall effect, etc \cite{zhang2005experimental,berger2006electronic, naumis2017electronic,pop2012thermal,allain2011klein,falomir2019optical,cea2012quantum,munoz2010ballistic,sang2019electronic}. Furthermore, due to the relativistic behavior of electrons, graphene provides a platform to study the  predicted  relativistic quantum mechanics effects, such as  Klein tunneling  \cite{young2009quantum,kim2017graphene,setare2019photonic,koke2020dirac,betancur2019electron,serna2019pseudospin}. Besides the great amount of theoretical studies,  the possibility of creating  different  two-dimensional  nanostructure has progressed  due to the rapid development in synthesis  and  artificial heterostructures growth techniques \cite{lin20162d,li2016heterostructures,csen2018one,farzaneh2019extrinsic}.\\
One of the main techniques  which has been established  is growing a
trilayer superlattice  in the (111) direction  for creating dice lattice ($\tau_3$ lattice) \cite{chakraverty2010controlled,okamoto2018transition,wang2011nearly}.
The dice lattice is  a honeycomb lattice with the addition of an extra atom is located  at the center of each hexagon which is described by the Dirac-Weyl Hamiltonian \cite{bhattacharya2019flat,manes2012existence,iurov2021tailoring,mandhour2020klein}.  The charge carriers at low energies in the dice lattice are described by the Weyl equation with pseudospin $S = 1$ . The gapless low-energy band structure of this lattice  consists of  Dirac cones and an additional flat band. This kind of band structure  has important consequences, such as magneto-optical conductivity, Hofstadter butterfly effect and zero-momentum optical conductivity\cite{illes2016magnetic, dey2020unconventional,chen2019nonlinear}. Furthermore, an attractive feature of dice lattice is that  it displays perfect transmission independent of the incident angle through a barrier, which is known as super-Klein tunneling (SKT) \cite{urban2011barrier,betancur2017super}. Recently many papers have been devoted to control transport property in this  lattice. Among the possibilities,  applying a time-dependent  potential can be  selected as an alternative approach to have  further control on particle transmission \cite{zhu2017fano,weekes2021generalized, li1999floquet, zhu2017shot} .\\
 Extrinsic additional sidebands at  energies $E\pm m\hbar \omega$ which correspond to Floquet channels $m=0,\pm 1, \pm 2,...$ induced by a time-periodic potential  have an important effect on transport properties \cite{li1999floquet,schulz2015scattering,zhu2017fano,cao2011massive,betancur2021anomalous,balassis2021temperature,bilitewski2015scattering}. These sidebands  arise in the context of  exchange of  energy quanta between  electrons and photons due to the oscillating potential. In fact, the  pseudospin-one particle can emit (absorb) photons and  drop (jump) to incident channels  or the other Floquet channels. Furthermore, the time-periodic potential can be created by applying a small ac signal in the setup pointed out in Ref.\cite{korniyenko2017shot}.  It is well known that  gap generation in the Dirac materials  is an important factor  in the development of  electronic devices \cite{gorbar2021gap,wehling2014dirac,xu2020anomalous,wang2017dirac,betancur2017perfect}. By band  gap engineering applications, we can open a gap in the energy dispersion of  gapless dice lattice. \\
In this paper, we briefly summarize the formalism used for a dice lattice in the presence of time-periodic potential. We study the transport properties  in the presence of an energy band gap  by considering various cases corresponding to different locations of the flat band. The aim of this kind of potential is that we can  have  good control of  the particles tunneling. Furthermore, 
the dice lattice has the  interesting property of being resistant to reflection of massive  pseudospin-one particle from the barrier at  energy $E=0.5 V_0$ \cite{urban2011barrier}. Here, we study and demonstrate the effect  of the time-periodic potential on the SKT. We use  analytical and numerical techniques to show the  dependence of transmission probability to the free parameters in our system.

\section{Theory}\label{Floquet}
Here, we study the effect of  time-periodic potential on transport property of pseudospin one Dirac-Weyl systems as shown in Fig.(\ref{sha1}). A tentative way for controlling the particle transmission is offered by increasing the number of degrees of freedom in our system. Thus, we study the transport of electrons  in the presence of a barrier with time-oscillating height. In this case, we can change the amplitude and frequency of the time-dependent potential to have further control on particle transmission. It is important to note that the time-square potential barrier causes arising many sidebands because of absorbing or emitting photons. We take into account the presence of a  small ac signal with potential  $V_{ac}$ and a static square potential $V_0$. Accordingly, the  potential is given by:
\begin{figure}[t]
    \centering
    \subfigure[]{\includegraphics[width=0.45\textwidth]{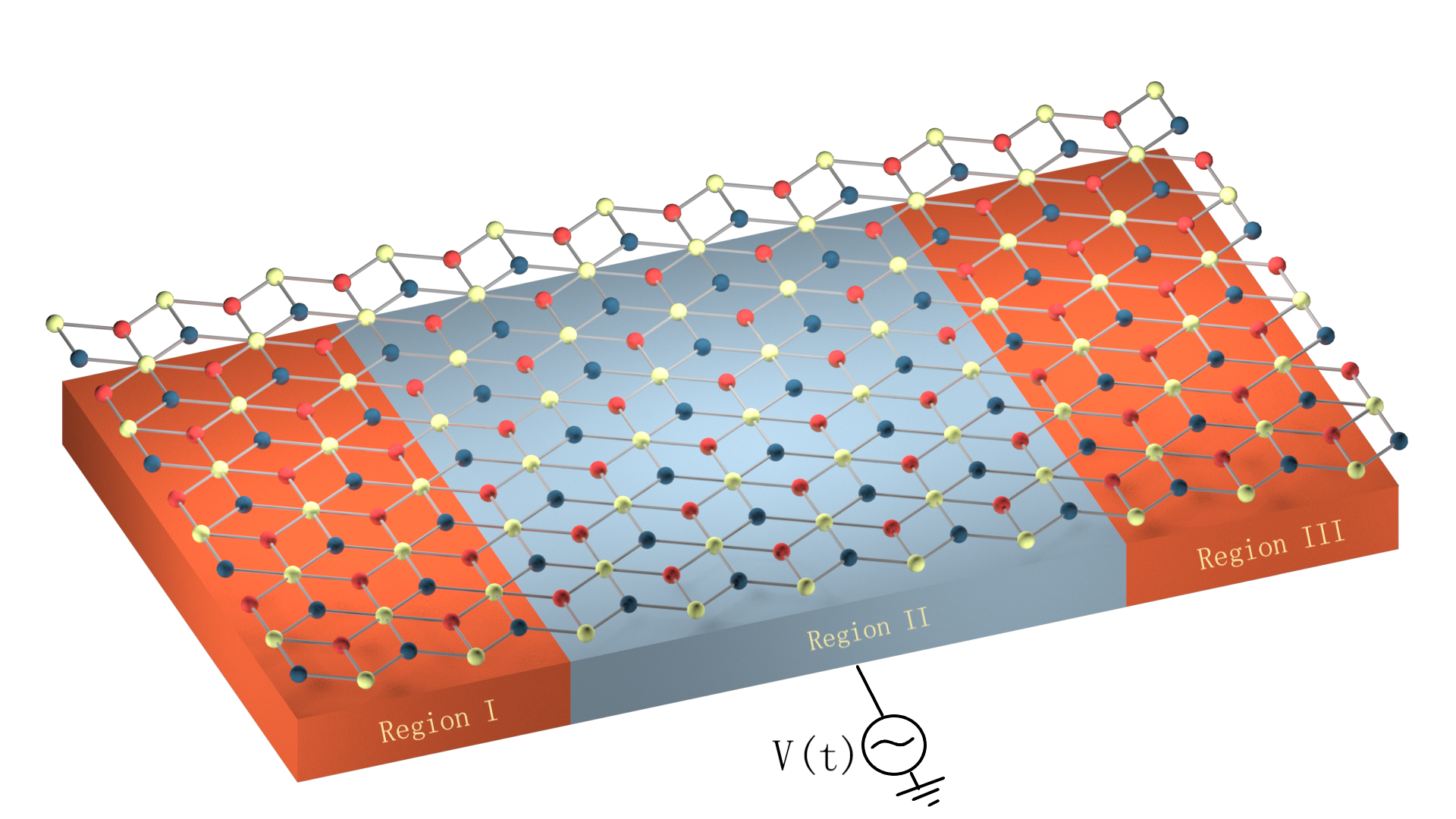}}\hspace{0.42em}%
    \subfigure[]{\includegraphics[width=0.45\textwidth]{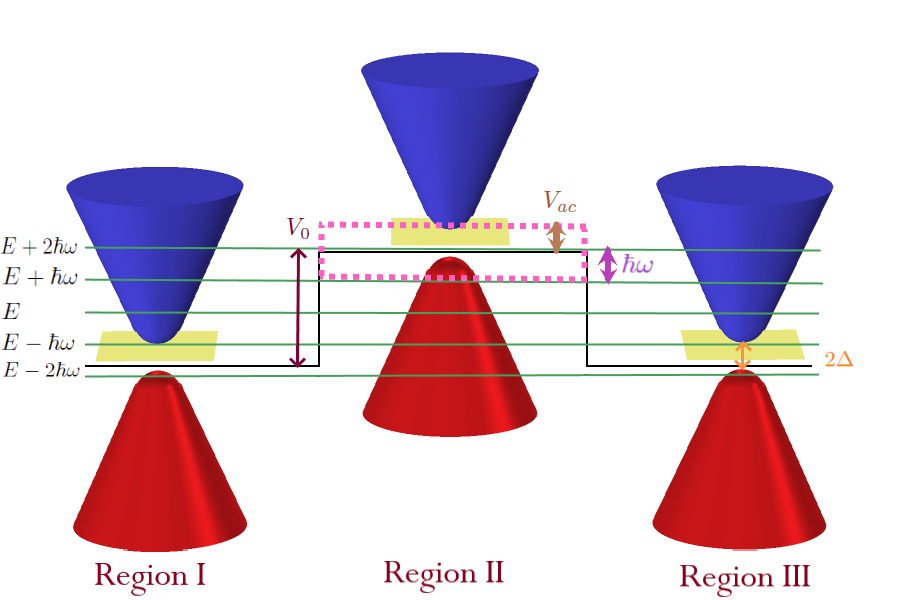}}
    \caption{ (a) The dice  lattice with three sites A,B and C ( denoted by red, blue, yellow color) in each unit cell in the presence of time-periodic potential. (b) Scheme of the band structure for the case $(l,L)= (1,1)$ across the time-periodic potential. The green lines indicate the  quantized energies $E \pm m\hbar \omega$.}
    \label{sha1}
\end{figure}

\begin{equation}\label{1}
V(x,t) =
\begin{cases}
V_0+V_{ac}\cos(\omega t)  \; \; \textrm{for} \; \; {0}<x< {D}, \\
0   \; \; \textrm{otherwise}.
\end{cases}
\end{equation}
Pseudospin-one particles in dice lattice in the presence of energy band gap $2\Delta$
 are governed by the following Hamiltonian,

\begin{equation}\label{Hsp}
H = v_F \vec{S}\cdot\vec{p} +  \Delta M+V(x,t),
\end{equation}
\noindent where $S = (Sx, Sy) $ is the x and y components of spin-one  matrices and $M$ describes the possible ways of  band gap opening which is given by \cite{betancur2017super}

\begin{eqnarray}
M = S_z = \left(\begin{array}{ccc} 
1 & 0 & 0\\
0 & 0 & 0\\
0 & 0 & -1
  \end{array}\right), & \quad & M = \pm U = \pm \left(\begin{array}{ccc} 
1 & 0 & 0\\
0 & -1 & 0\\
0 & 0 & 1
  \end{array}\right). \nonumber\\
& 
\end{eqnarray}

\noindent It is noteworthy that $M=S_z$  corresponds to the flat band  at the the middle of the band gap
which is  denoting by $L=0,l=0$ for inside and outside the barrier  respectively.
 Furthermore, the case where the flat band is located at the bottom (top) of the
conduction (valence) band is described by  $M=+U$ ($M=-U$)  with
with $l = 1 (l = -1)$ for the outside and $L = 1 (L = -1)$ for the inside the potential barrier.  The solution to the wave equation $(H+V(x,t))\Phi(r,t)=i\hbar \partial_t \Phi(r,t) $, which acts on three-component spinor   $\Phi(\vec{r,t}) = \textrm{e}^{ik_yy}(\Phi_1(x,t),\Phi_2(x,t),\Phi_3(x,t))$, inside the barrier (region II) can be written as

\begin{equation}\label{2}
\Phi^\textrm{II}(r,t)= \frac{e^{i k_y y}e^{-i Et/\hbar}}{2}\sum_{n,m=-\infty}^{\infty}J_{n}(\alpha) \Bigg[\begin{pmatrix}
a'_{L,m} e^{-i{\xi}_m}  \\
 s'_mb'_{L,m}\\
c'_{L,m}e^{i{\xi}_m}
\end{pmatrix}t'_m e^{iq_{x,m}x}+
\begin{pmatrix}
a'_{L,m} e^{i{\xi}_m}\\
-s'_mb'_{L,m } \\
  c'_{L,m} e^{-i{\xi}_m}
\end{pmatrix}r'_m e^{-iq_{x,m}x}\Bigg]e^{-i(n+m)\omega t},
\end{equation}
\noindent where $\alpha=V_{ac}/ \hbar \omega $ and 
the phases of pseudospin for  sideband $m=0,\pm 1, \pm 2,...$ is $\xi_m=\arctan(k_y/q_{x,m})$. On the other hand the factors $a'_{L,m}$, $b'_{L,m}$ and $c'_{L,m}$ in this region, are defined as
\begin{subequations}
\begin{eqnarray}
a'_{L,m} & = & L^2\sqrt{1 + \frac{L \Delta'}{E-V_0-m \hbar \omega }} + (1 - L^2)\left(1+\frac{\Delta'}{E-V_0-m \hbar \omega }\right),\\
b'_{L,m} & = & L^2\sqrt{2\left(1 - \frac{ L \Delta'}{E-V_0-m \hbar \omega }\right)}+(1 - L^2)\sqrt{2\left(1-\frac{\Delta'^2}{(E-V_0-m \hbar \omega )^2}\right)},\\
c'_{L,m} & = & L^2a'_l+(1 - L^2)\left(1-\frac{\Delta'}{E-V_0-m \hbar \omega }\right).
\end{eqnarray}
\end{subequations}
\noindent The corresponding wave functions in the incident and transmitted sides, can be expressed as:

\begin{equation}\label{3}
\Phi^\textrm{I}(r,t)= \frac{e^{i k_y y}e^{-i Et/\hbar}}{2}\sum_{m=-\infty}^{\infty} \Bigg[\begin{pmatrix}
a_{l,m} e^{-i{\phi}_m}  \\
 s_mb_{l,m}\\
c_{l,m }e^{i{\phi}_m}
\end{pmatrix}e^{ik_{x,m}x} \delta_{m,0}+
\begin{pmatrix}
a_{l,m} e^{i{\phi}_m}\\
-s_mb_{l,m} \\
c_{l,m} e^{-i{\phi}_m}
\end{pmatrix}r_m e^{-ik_{x,m}x}\Bigg]e^{-im\omega t},
\end{equation}
and

\begin{equation}\label{4}
\Phi^\textrm{III}(r,t)= \frac{e^{i k_y y}e^{-i Et/\hbar}}{2}\sum_{m=-\infty}^{\infty} \Bigg[
\begin{pmatrix}
a_{l,m} e^{-i{\phi}_m}\\
s_mb_{l,m} \\
  c_{l,m}e^{i{\phi}_m}
\end{pmatrix}t_m e^{i k_{x,m} x}\Bigg]e^{-im\omega t},
\end{equation}

\noindent where $\phi_0$ is the angle of incidence and  $\phi_m=\arctan(k_y/k_{x,m})$. Here, the band indices $s_m=\sgn(E-m\hbar\omega)$ and $s_m^\prime=\sgn(E-V_0 -m\hbar\omega)$  with the positive (negative) sign  indicate the conduction (valence) band for the inside and outside the barrier, respectively. In the outside of barrier, the coefficients $a_{l,m},b_{l,m},c_{l,m}$ are given by:

\begin{subequations}
\begin{eqnarray}
a_{l,m} & = & l^2\sqrt{1 + \frac{l \Delta}{E-m \hbar \omega }} + (1 - l^2)\left(1+\frac{\Delta}{E-m \hbar \omega}\right),\\
b_{l,m} & = & l^2\sqrt{2\left(1 - \frac{l \Delta}{E-m \hbar \omega}\right)}+(1 - l^2)\sqrt{2\left(1-\frac{\Delta^2}{(E-m \hbar \omega)^2}\right)},\\
c_{l,m} & = & l^2a_l+(1 - l^2)\left(1-\frac{\Delta}{E-m \hbar \omega}\right).
\end{eqnarray}
\end{subequations}
The linear momentum in  $y$ direction is conserved as the  Hamiltonian commutes with it  and can be expressed as 
\begin{equation}
  k_{0,y}  =  \frac{\sqrt{E^2 - \Delta^2}}{\hbar v_F}\sin\phi.
\end{equation}
On the other hand, the $x$ components of the wave vectors for the outside and inside the barrier, respectively is given by:

\begin{subequations}
\begin{eqnarray}
k_{x,m} =  \sqrt{\frac{(E - m\hbar\omega)^2 - \Delta^2}{\hbar^2v_F^2}-{k_{0,y} ^2}}, \\
q_{m,x} = \sqrt{\frac{(E-V_0-m\hbar\omega)^2 - \Delta'^2}{
\hbar^2v_F^2}-{k_{0,y} ^2}}.
\end{eqnarray}
\end{subequations}
\noindent According to the above expressions, wave vectors in  the time-periodic potential can be imaginary and  consequently the wave function becomes evanescent\cite{ szabo2013relativistic,atteia2017ballistic}. For  grazing incidence angles and largest sidebands, the evanescent modes  appear in the barrier.  We   take these evanescent modes into account in the  numerical result. By  using  the boundary conditions at $x=0$, $x=D$ in which   $\Phi_1(x,t)+\Phi_3(x,t)$ and $\Phi_2(x,t)$ must be continuous, we can  find the reflected $(r_m)$ and transmitted $(t_m)$ amplitudes in the following system of equations

\begin{widetext}

\begin{figure}[t]
    \centering
    \subfigure[]{\includegraphics[width=0.4\textwidth]{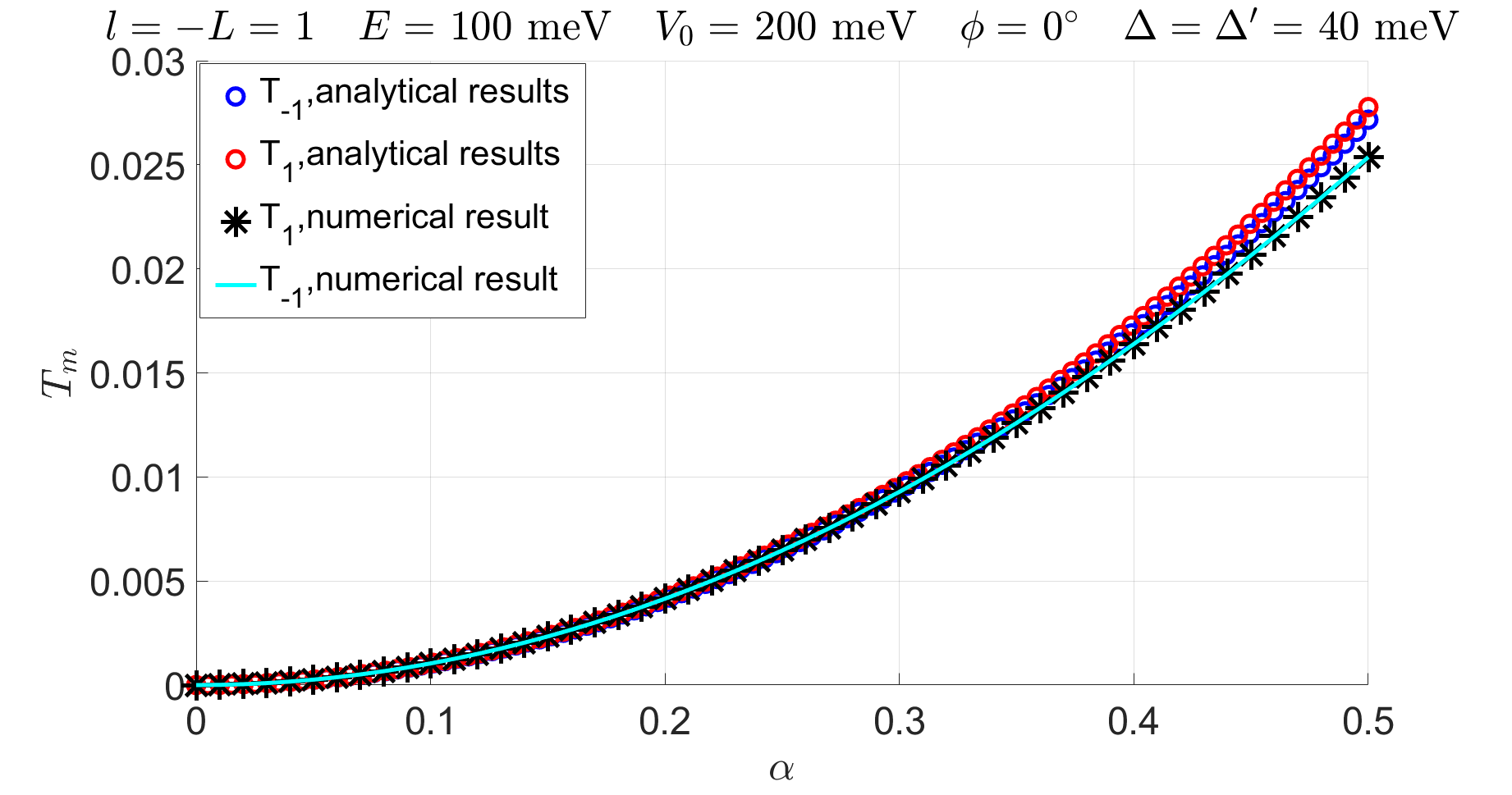}}\hspace{0.1em}%
    \subfigure[]{\includegraphics[width=0.4\textwidth]{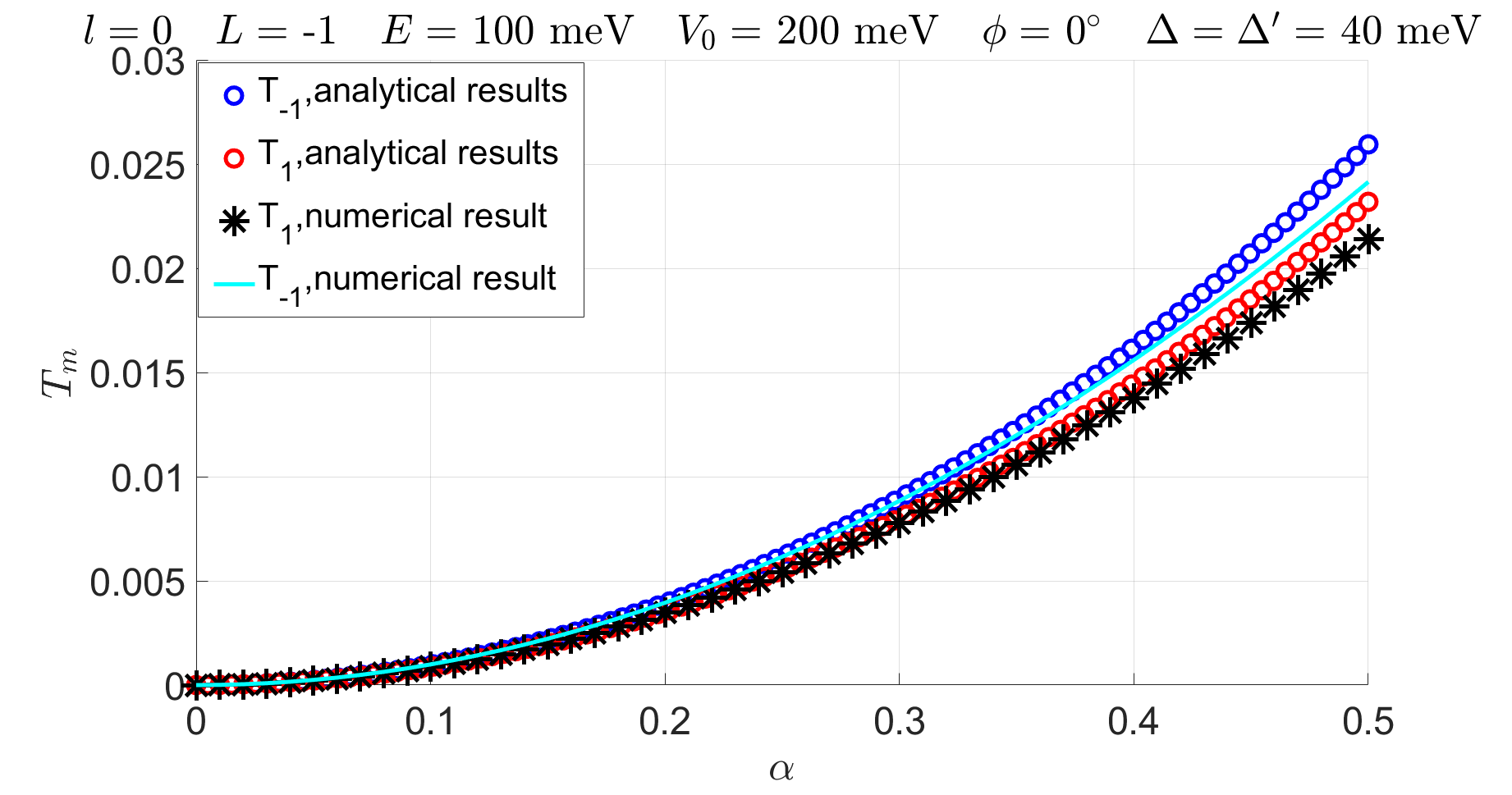}}
    \subfigure[]{\includegraphics[width=0.4\textwidth]{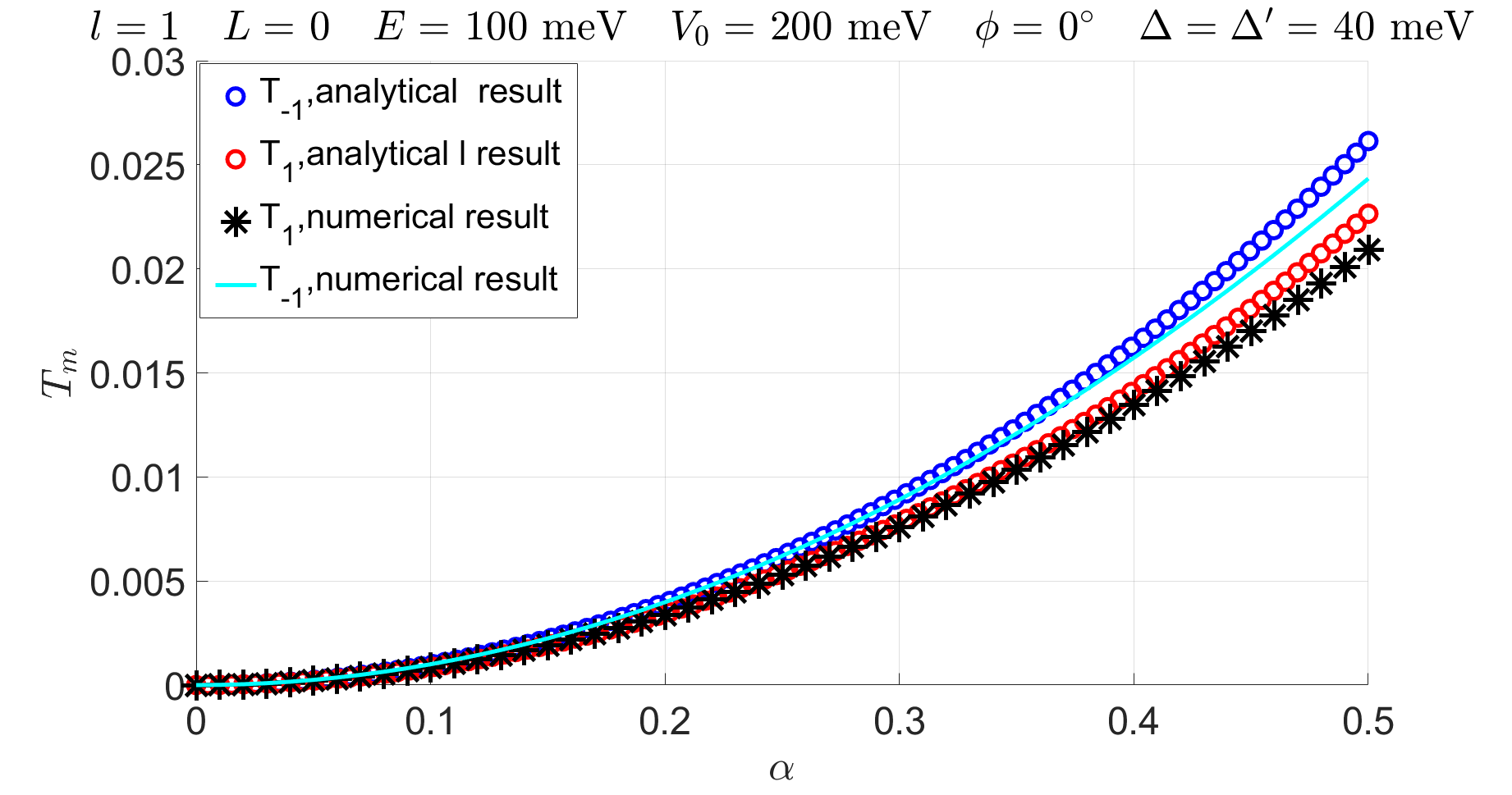}} \hspace{0.1em}%
    \subfigure[]{\includegraphics[width=0.4\textwidth]{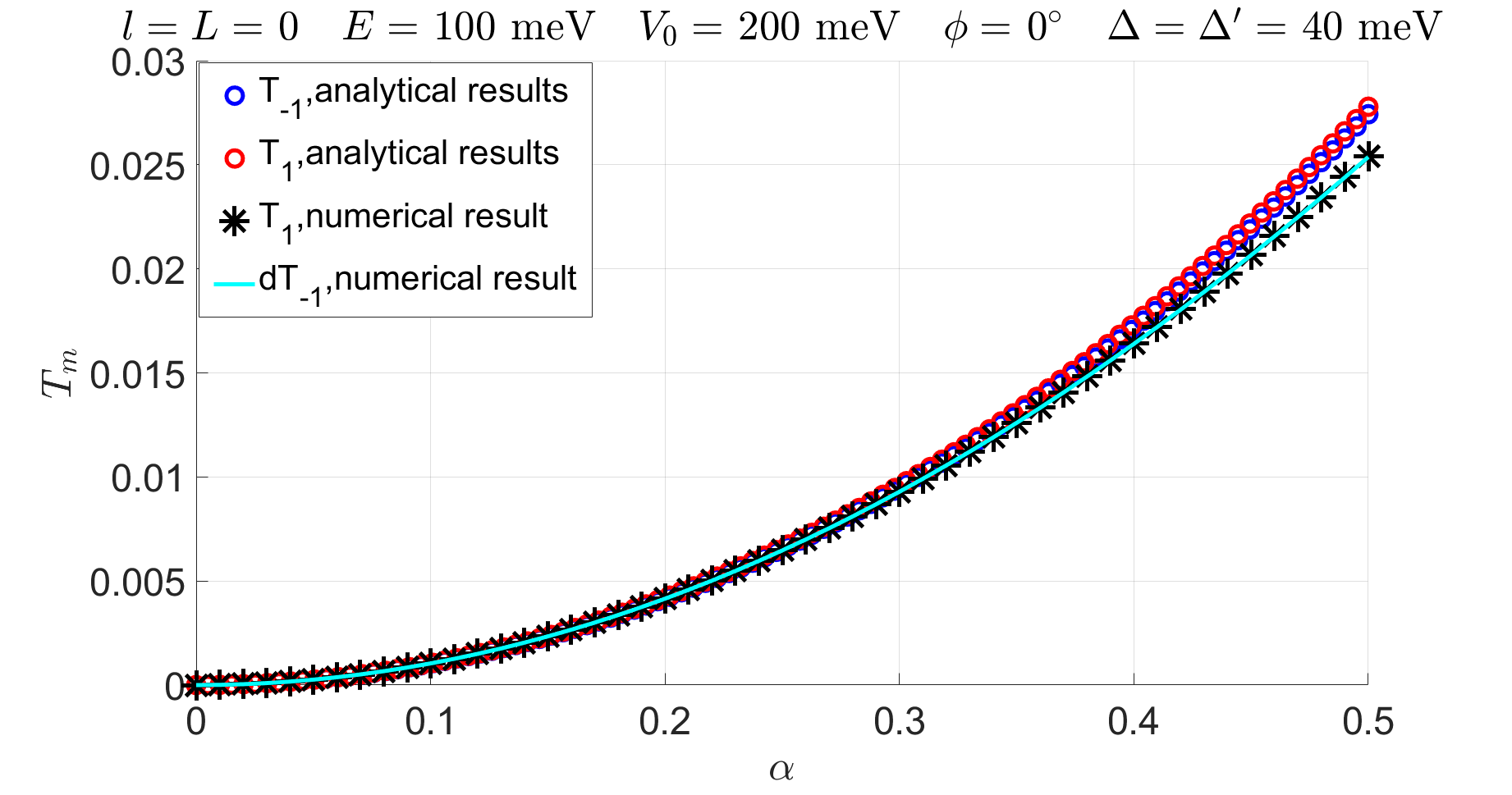}}
    \caption{ The transmission coefficient for first  sidebands as a function of $\alpha$ for normal incident wave with $\phi_0=0$ where $E=100$ meV and $V_0=200$ meV and $D=100$ nm. The open symbols  represent the  the analytical expression (\ref{8}) for first sidebands and cyan curve and stars symbols are the numerical results  considering  Floquet  sidebands   $m=0,\pm1 , ...,\pm12$.(a) $(1,-1)$, (b) $(0,-1)$ (c) $(1,0)$(d) $(0,0)$. }
    \label{shg}
\end{figure}

\end{widetext}

\begin{subequations}
\begin{eqnarray}\label{ta33}
s_0 b_{l,0} - s_m b_{l,m} r_m  &=&  \sum_{n=-\infty}^{\infty}s'_n b^{'}_{L,n}(t^{'}_n - r^{'}_n)J_{m-n}(\alpha),\\
a_{l,0} e^{-i\phi_0} +  c_{l,0} e^{i\phi_0}+[a_{l,m} e^{i\phi_m} +  c_{l,m} e^{-i{\phi}_m}]r_m & =
& \sum_{n=-\infty}^{\infty}\bigg\{[a^{'}_{L,n} e^{-i{\xi}_n} + \nonumber\\
&& \qquad c^{'}_{L,n }e^{i{\xi}_n}]t^{'}_n + [a^{'}_{L,n} e^{i{\xi}_n} +  c^{'}_{L,n} e^{-i{\xi}_n }]r^{'}_n \bigg\}J_{m-n}(\alpha), \\
s_mb_{l,m} t_m e^{i k_{x,m} D}&=&\sum_{n=-\infty}^{\infty} J_{m-n}({\alpha}) \left[t^{'}_n e^{iq_{x,n}D}-r^{'}_n
e^{-iq_{x,n} D}\right]s^{'}_n b^{'}_{L,n} ,\\
(a_{l,m} e^{-i{\phi}_m} +  c_{l,m} e^{i{\phi}_m})t_m e^{i k_{x,m} D} & = & \sum_{n=-\infty}^{\infty}J_{m-n}({\alpha})
\bigg\{ [a^{'}_{L,n} e^{-i{\xi}_n } +  c^{'}_{L,n} e^{i{\xi}_n }]t^{'}_n e^{iq_{x,n}D} + \nonumber\\
 && [a^{'}_{L,n} e^{i{\xi}_n} +  c^{'}_{L,n} e^{-i{\xi}_n }]r^{'}_n e^{-iq_{x,n} D}\bigg\}.
\end{eqnarray}
\end{subequations}
\noindent To solve the above infinite number of coupled
equations, one approximation must be used. For small value of $\alpha$, we can  consider limited sidebands since Bessel functions are small for large value of $m$. By   using a finite  Floquet sidebands  with $m=0, \pm 1,..., \pm u$ and by writing the above  system of equations 
in a matrix form  $\Upsilon X=\mu$, we can obtain the transmission coefficient.  Here, $\Upsilon, X$ have the following  forms:

\begin{figure}[t]
    \centering
    \subfigure[]{\includegraphics[width=0.48\textwidth]{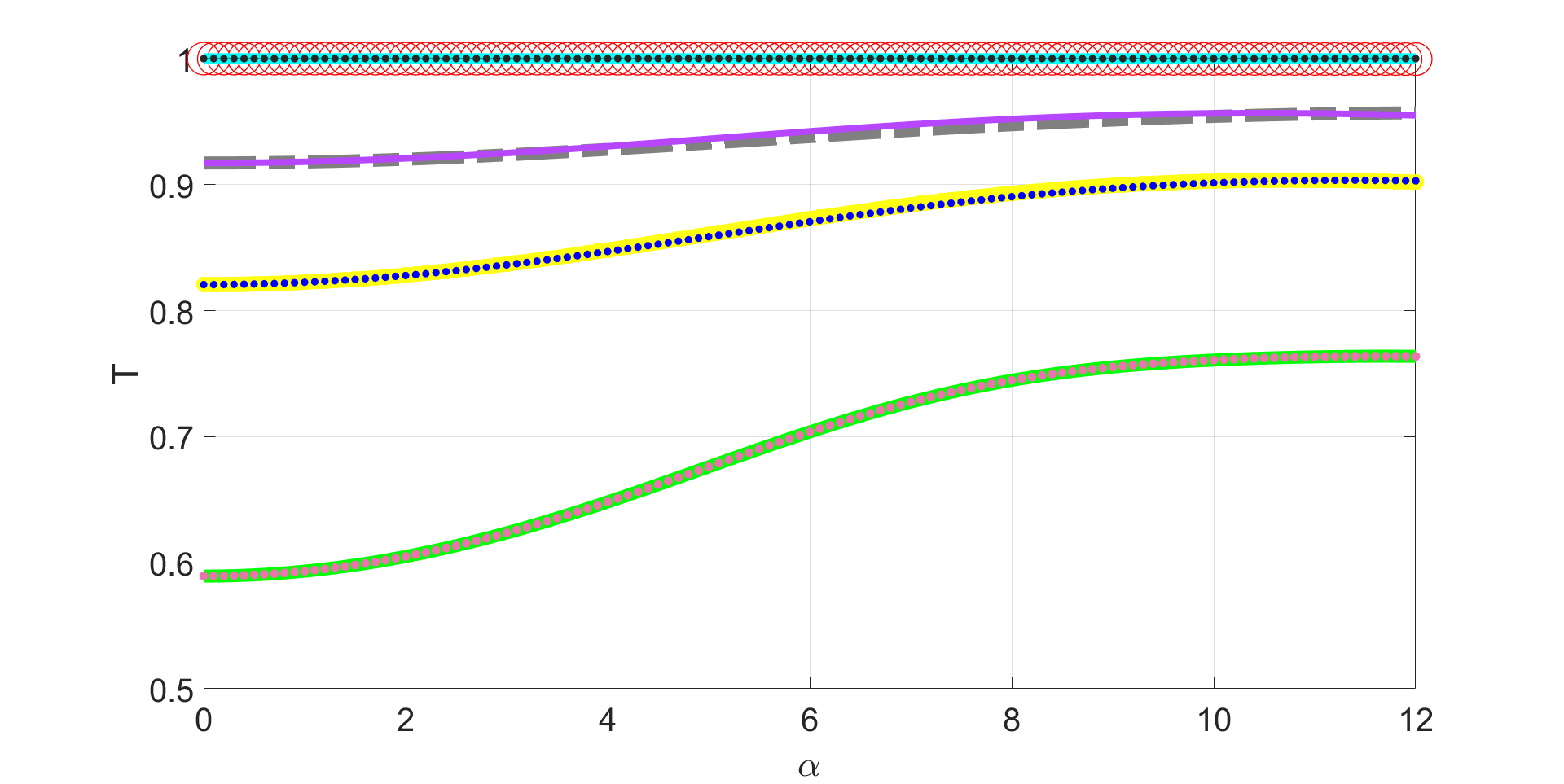}}\hspace{0.1em}%
    \subfigure[]{\includegraphics[width=0.48\textwidth]{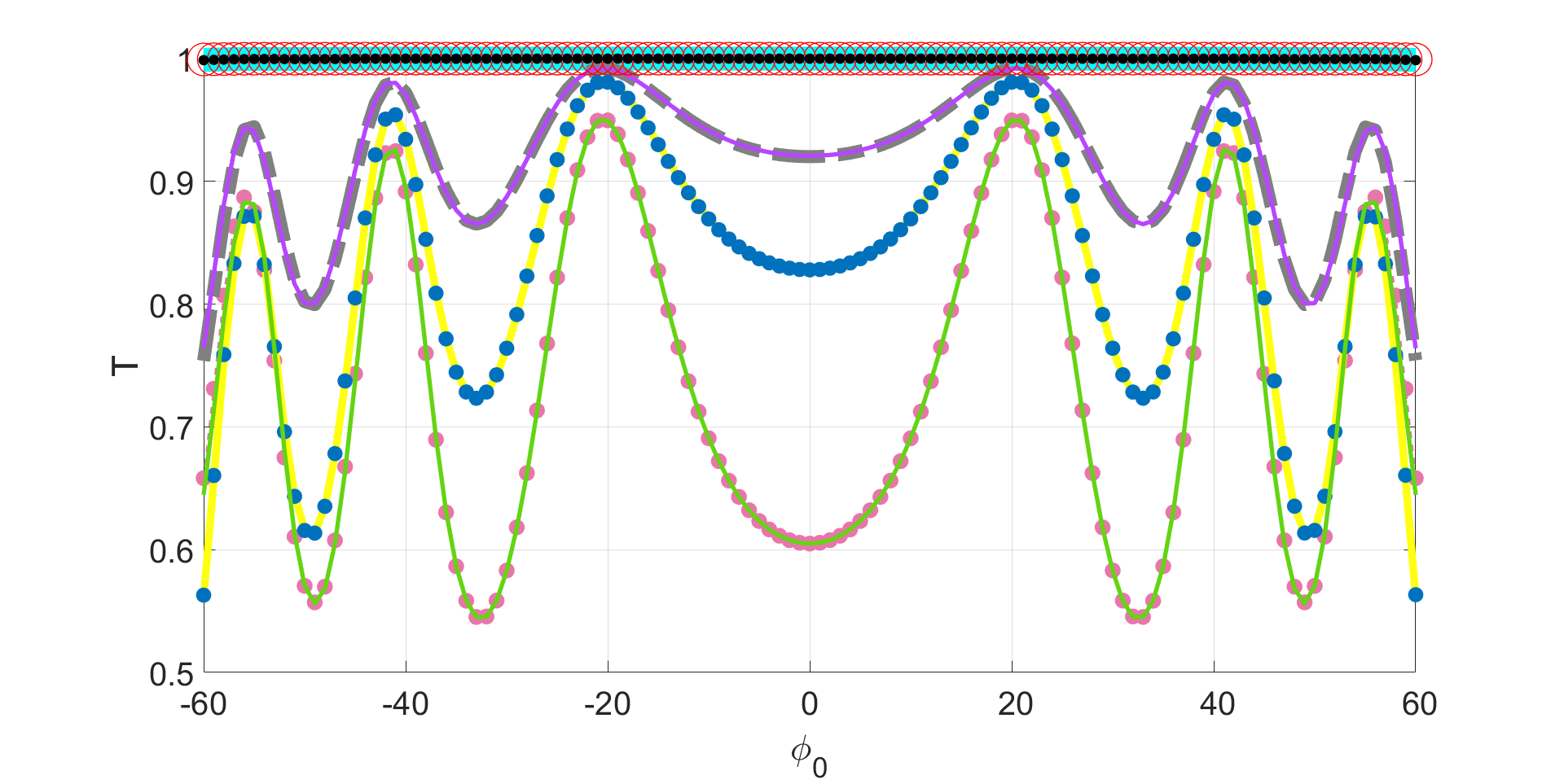}}
    \subfigure[]{\includegraphics[width=0.48\textwidth]{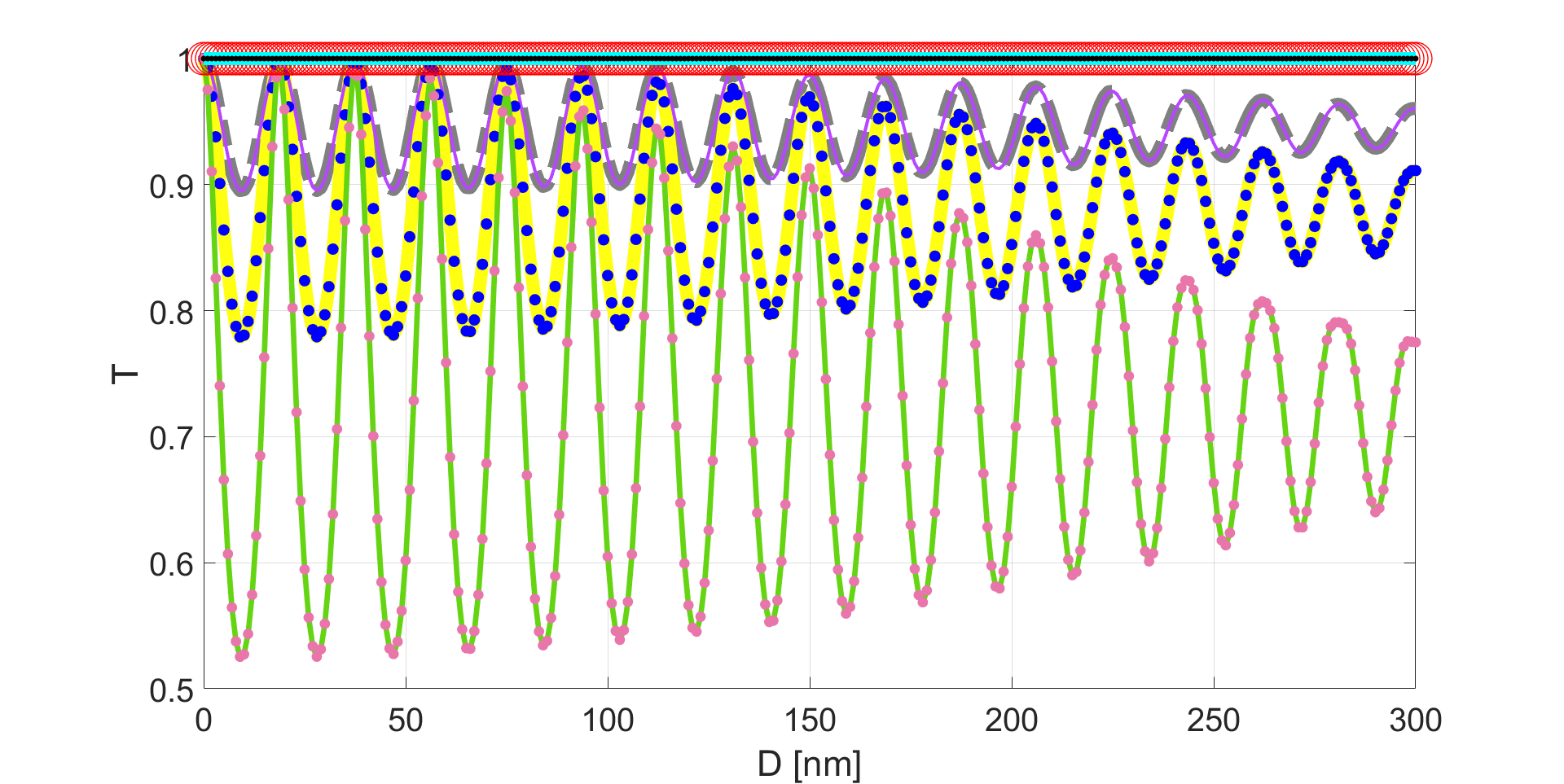}}\hspace{0.1em}
    \subfigure[]{\includegraphics[width=0.47\textwidth]{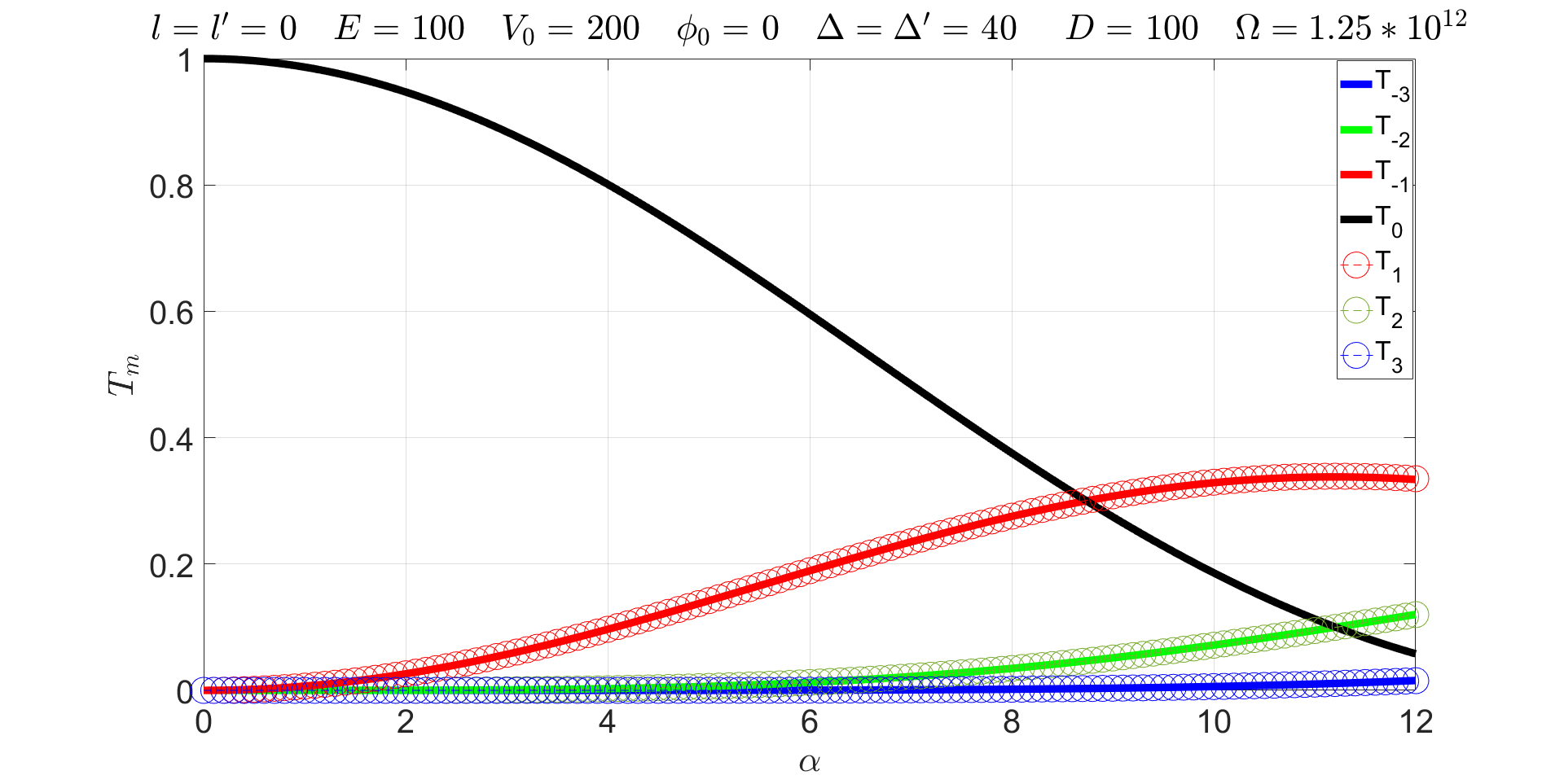}}
    \caption{   The total transmission as functions of:(a) $ \alpha$ in  normal incidence and $D=100$ (b) incidence angle with   $D=100$, $ \alpha=2$ and (c) width of  the barrier for the following cases $(l,L)$: $(1,-1)$ (red-circle), $(-1,1)$ (black-dot), $(0,0)$  (cyan-solid line),
      $(0,1)$ (yellow-solid line), $(-1,0)$ (blue-dot),
     $(0,-1)$  (purple-solid line)  ,$(1,0)$ (gray-dashed line), $(-1,-1)$ (pink-dot),     and $(1,1)$ (green-solid line).
    (d) Transmission probability $T_m = |t_m|^2$ with $m=-3,..,3$ for the case (0,0) with energy $E=V_0/2$ and $\Delta=\Delta^\prime$. }
    \label{shga}
\end{figure}

\begin{equation}\label{num1}
\Upsilon=[ \Upsilon_1 \Lambda_1 \quad   \Upsilon_1\Lambda_2 \quad \Upsilon_2\Lambda_3 D \quad \Upsilon_2\Lambda_2D]^T,
\end{equation}
and
\begin{equation}\label{num2}
X=(r_{-u},...,r_{u},r'_{-u},...,r'_{u},t'_{-u},...,t'_{u}, t_{-u},...,t_{u})^T,
\end{equation}
with ${\Upsilon}_1 = [-\mathcal{I} \quad \mathcal{J} \quad \mathcal{J} \quad \mathcal{O}]$  and $\Upsilon_2 = [\mathcal{O} \quad \mathcal{J} \quad\mathcal{J}\quad -\mathcal{I}]$. We use the short notations:

\begin{eqnarray}
\mathcal{J} & = & {\mathcal J}_{mn} = J_{m -n}(\alpha),\nonumber\\
{\mathcal \mu} &=& [\mu_1 \quad\mu_2 \quad \mathcal{O} \quad\mathcal{O}]^T,\nonumber\\
{ \Lambda_1}&=& \diag\left[\left(-s_m b_m\right)_{m=-u}^u \left(-s'_m b'_m\right)_{m=-u}^u \left(s'_m b'_m\right)_{m=-u}^u \left(s_m b_m\right)_{m=-u}^u\right],
\nonumber\\
{ \Lambda_2}&=& \diag\left[\left(a_m e^{i \phi_m}+c_m e^{-i \phi_m}\right)_{m=-u}^u \left(a'_m e^{i \xi_m}+c'_m e^{-i \xi_m}\right)_{m=-u}^u\left(a'_m e^{-i \xi_m}+c'_m e^{i \xi_m}\right)_{m=-u}^u \right. \nonumber\\
 && \qquad \qquad \left. \left(a_m e^{-i \phi_m}+c_m e^{i \phi_m}\right)_{m=-u}^u\right],
\nonumber\\
{ \Lambda_3}&=& \diag\left[\left(s_m b_m\right)_{m=-u}^u \left(-s'_m b'_m\right)_{m=-u}^u \left(s'_m b'_m\right)_{m=-u}^u\left(s_m b_m\right)_{m=-u}^u\right],
 \end{eqnarray}
 and define

\afterpage{
\begin{figure}[t]
    \centering
    \subfigure[]{\includegraphics[width=0.4\textwidth]{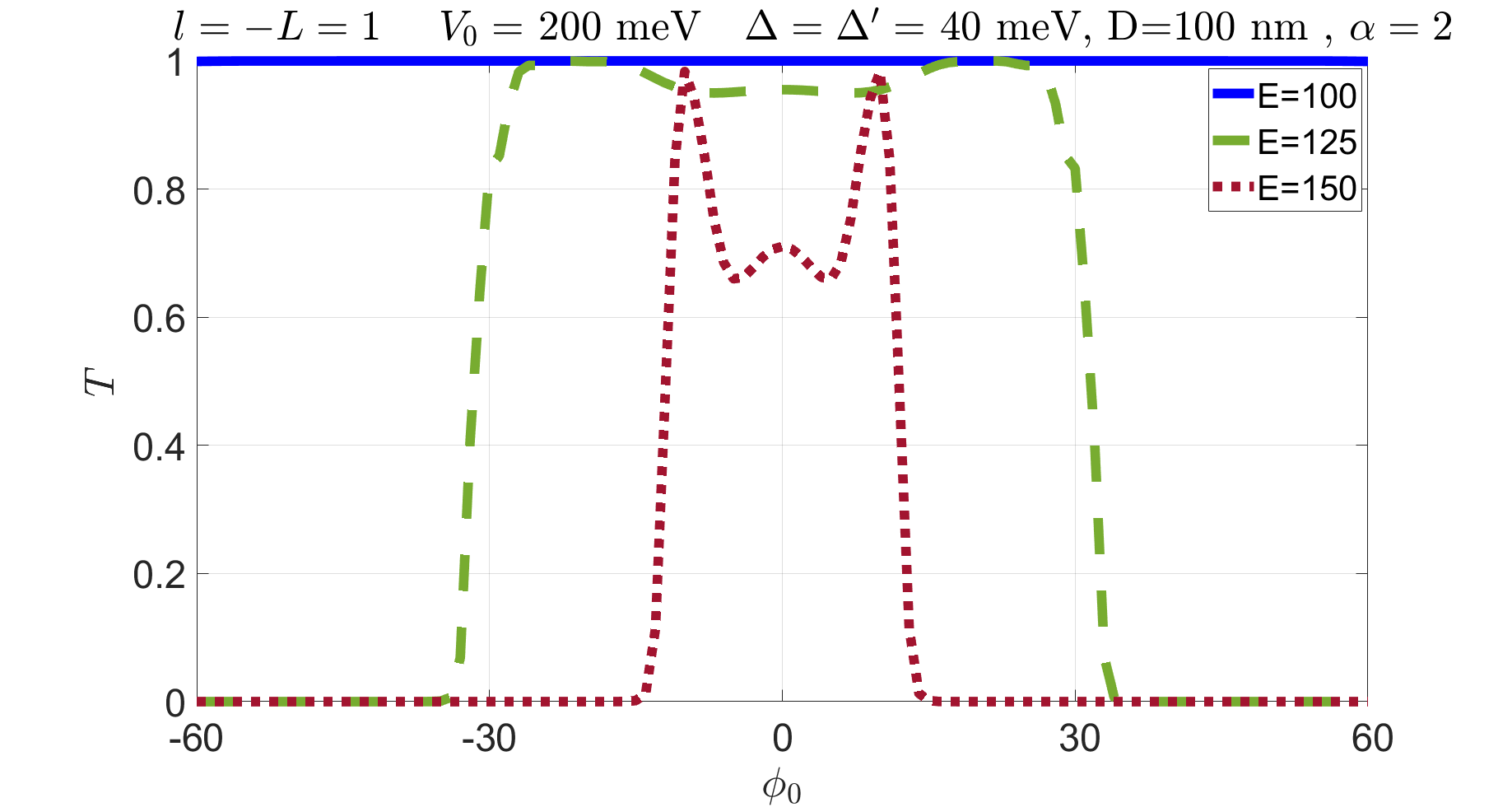}}\hspace{0.01em}%
    \subfigure[]{\includegraphics[width=0.4\textwidth]{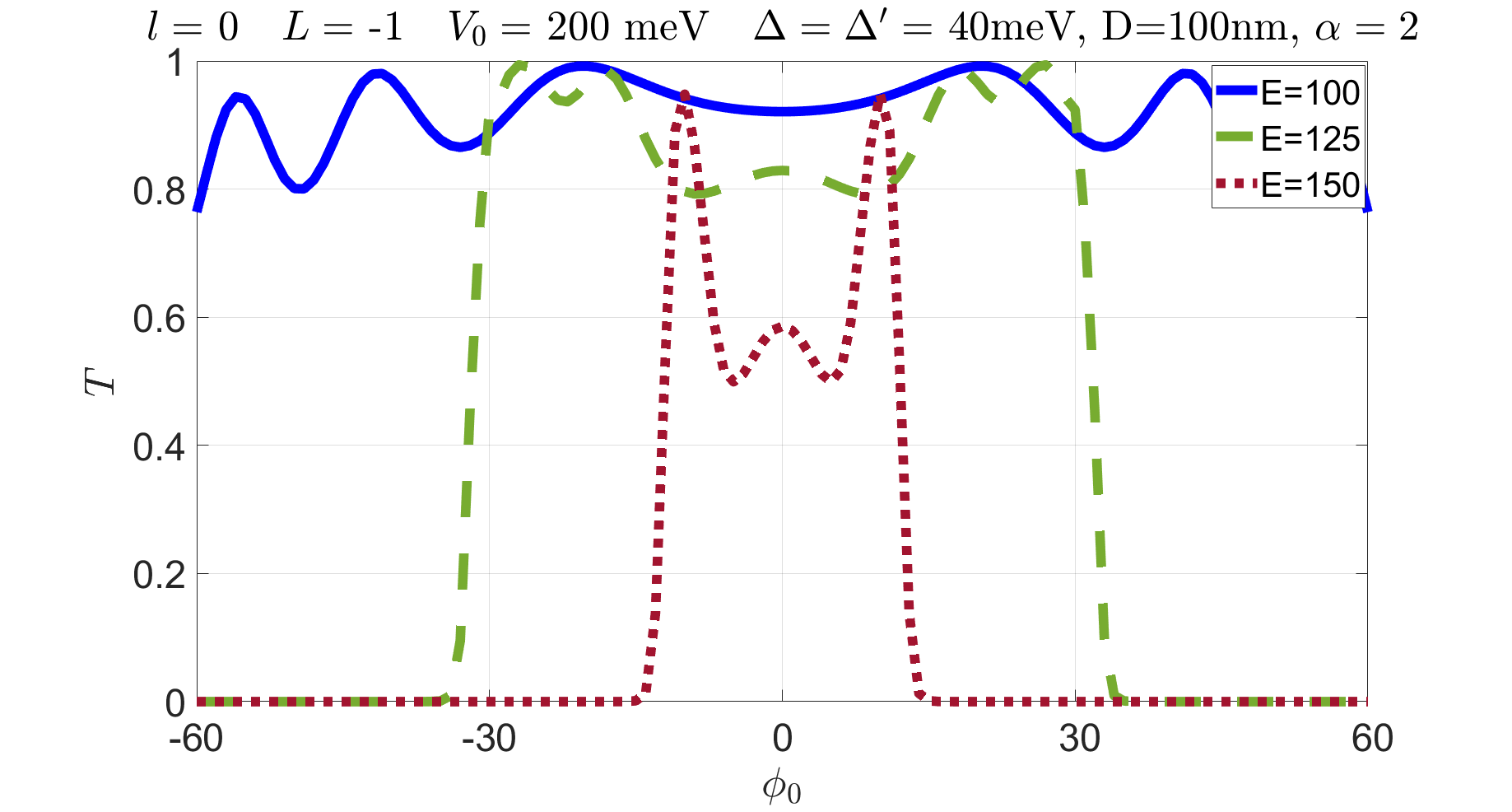}}
    \subfigure[]{\includegraphics[width=0.33\textwidth]{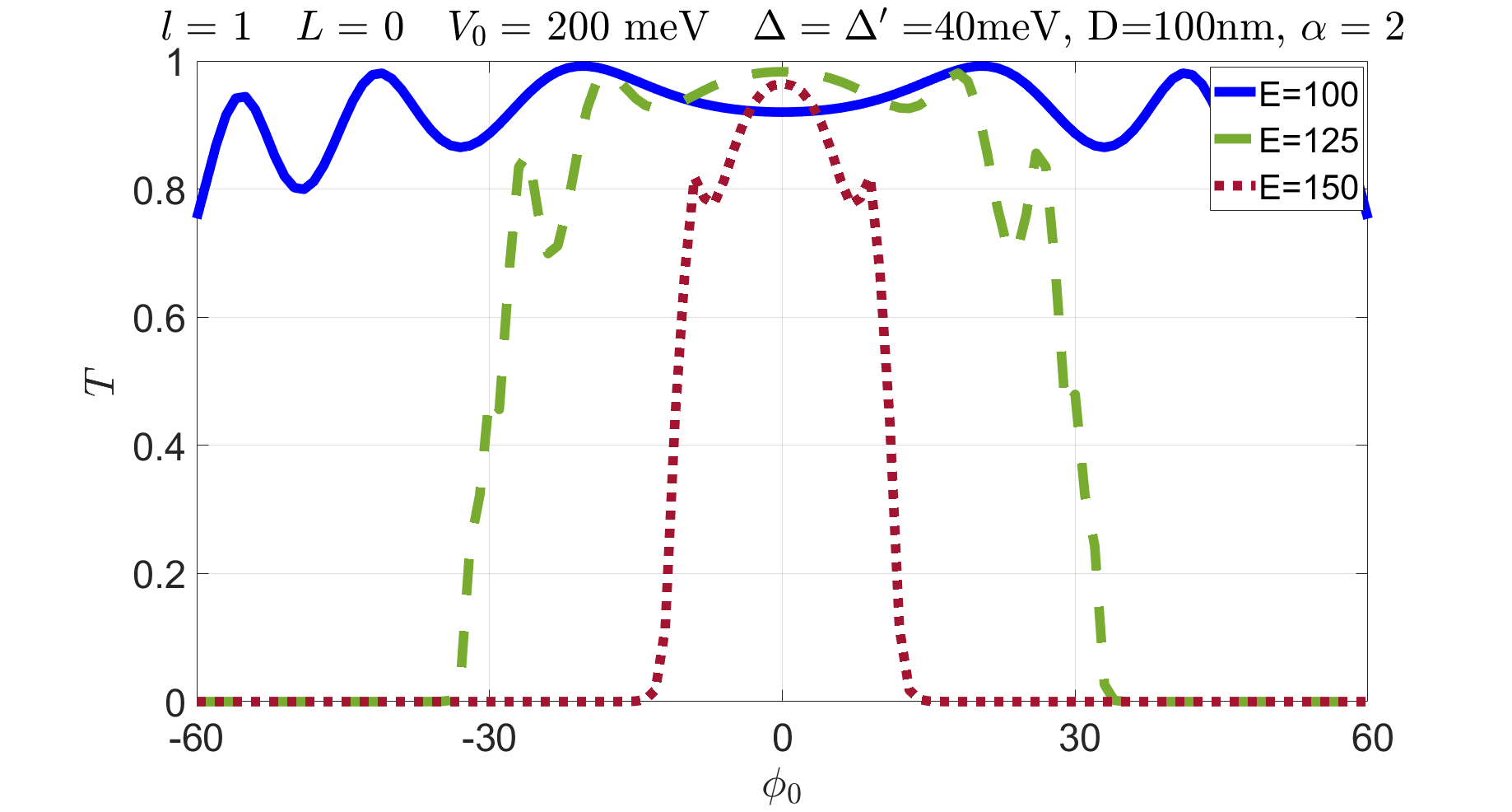}}\hspace{0.01em}%
    \subfigure[]{\includegraphics[width=0.33\textwidth]{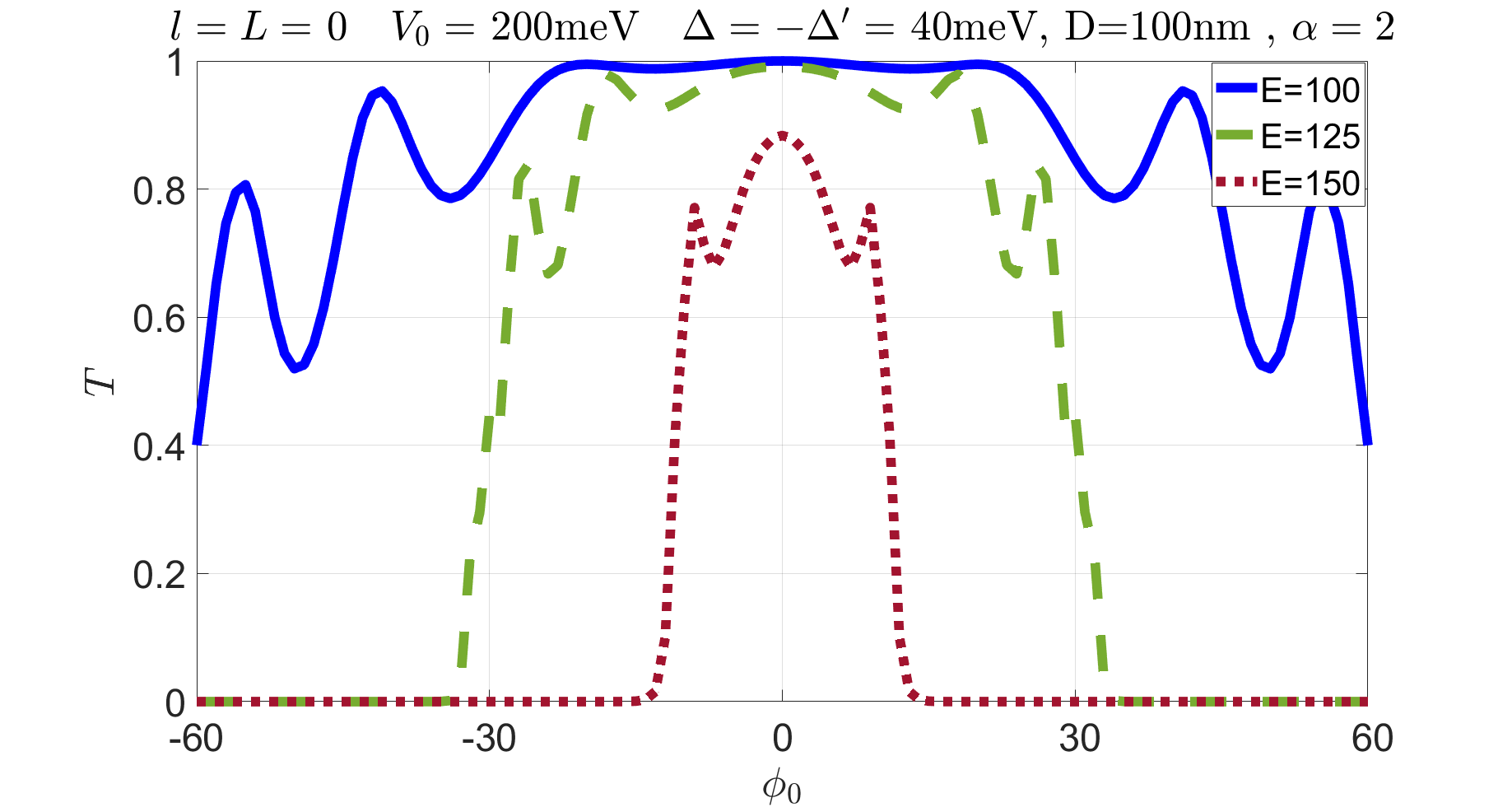}}
    \subfigure[]{\includegraphics[width=0.33\textwidth]{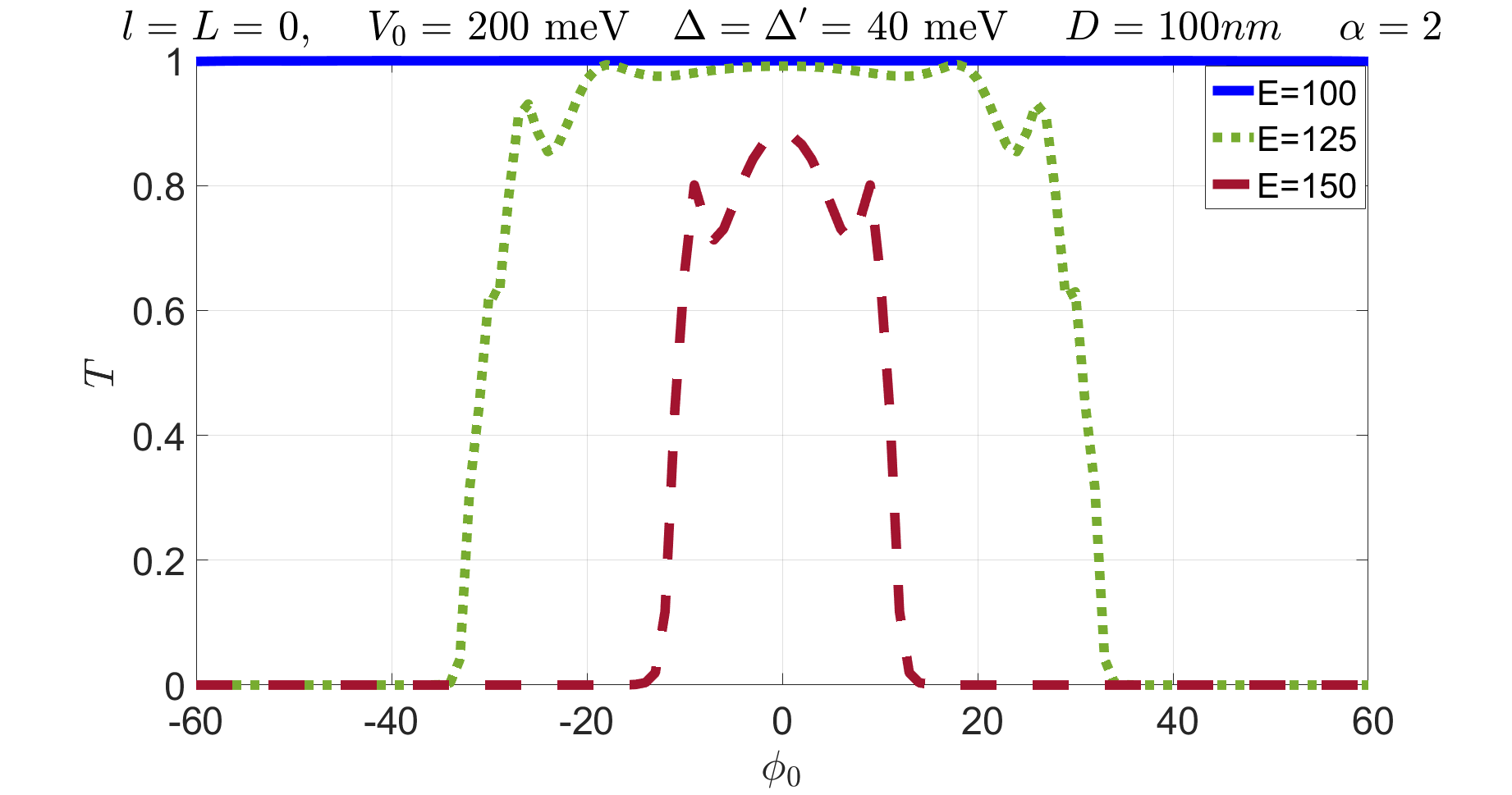}}
    \caption{ The angular dependence of the total transmission  through a barrier of width $D=100nm$ at various energies with  $V_0=200$ meV, $\alpha=2$ and $\Omega=1.25\times 10^{12}$Hz by considering $-12\leq m \leq 12$ for different cases: (a) $(1,-1)$(b) $(0,-1)$ (c) $(1,0)$ (d) $(0,0)$ with $\Delta=-\Delta^ \prime$ (e) $(0,0)$ with $\Delta=\Delta^ \prime$. }
    \label{sha3}
\end{figure}
 \begin{table}[b]
  \centering
    \begin{tabular}{ | l | l | l | l | p{1.5cm} |}
    \hline
   notation& formula  \\ \hline

     $Z_j$       & $b_{l,j} \Omega_0^{*}-{b_{L,0}^{\prime}}\gamma_j$    \\ \hline
    $W_j$         & $b_{l,j} \Omega_0+{b_{L,0}^{\prime}}\gamma_j$   \\ \hline
    $F_j$          & $b_{l,j} \Omega_j^{*}-{b_{L,j}^{\prime}}\gamma_j$   \\ \hline
    $N_j$          &  $b_{l,j} \Omega_j+{b_{L,j}^{\prime} }\gamma_j$   \\ \hline
    \hline
    \end{tabular}
\caption{The set of used  parameters in  Eq.(\ref{8}).}
\label{ta3}
\end{table}}

\begin{eqnarray}
{D}&=&
\diag\left[\left(e^{-ik_{x,m}D}\right)_{m=-u}^u \left(e^{-iq_{x,m}D}\right)_{m=-u}^u \left(e^{iq_{x,m}D}\right)_{m=-u}^u \left(e^{ik_{x,m}D}\right)_{m=-u}^u\right],\nonumber\\
\vec \mu_1&=&\big(s_m b_m
\delta_{m,0}
\big)_{m=-u}^u\ ,\nonumber\\
\vec{\mu}_2&=& \big((a_m e^{-i\phi_m}+c_m e^{i\phi_m})\delta_{m,0}
\big)_{m=-u}^u 
\end{eqnarray}
\noindent The  transmission coefficient $T_m=|t_m|^2$ of the  scattering problem can be obtained  analytically for the first  sidebands.  In the limiting case when the parameter $ \alpha $ is small, we can obtain  the  transmission coefficient  $t_j$  (with $j=\pm 1$) for the first sidebands, as following:

\begin{widetext}

\begin{multline}\label{8}
t_j={\frac{e^{-ik_jD}e^{ik_0D}t_0 b_{l,0} J_j(\alpha)}{b_{l,j} b_{L,0}^{\prime}(F_0^{*}+W_0^{*})J_0(\alpha)}}
\Bigg( F_0^{*}(b_{L,0}^{\prime}+\frac{b_{L,j}^{\prime}W^{*}_j}{F^{*}_j})+W_0^{*}(b_{L,0}^{\prime}-\frac{b_{L,j}^{\prime}Z_j^{*}}{F_j^{*}}) \\
-(\frac{e^{iq_jD}b_{L,j}^{\prime}(F_j^{*}+N_j^{*})}{(| F_j|^2 -| N_j|^2 e^{2i q_j D})})\Big[F_0^{*}Z_j e^{-iq_0D}-W_0^{*}W_je^{iq_0D}+\frac{{N_j}{e^{iq_jD}}(Z_j^{*}W_0^{*}-F_0^{*}W_j^{*})}{F_j^{*}}\Big]
\Bigg).
\end{multline}
\end{widetext}
\noindent In order to simplify the notations, we use  variables which are summarized in table (\ref{ta3}) and the following  definitions:

\begin{subequations}
\begin{eqnarray}
\gamma_j & = & l^2 a_{l,j} \cos(\phi_j)+(1-l^2) [i a_{l,j} \sin(\phi_j)+e^{-i\phi_j}],\\
\Omega_j & = & L^2  {a_{L,j}^{\prime}} \cos(\xi_j)
+(1 - L^2)[{i a_{L,j}^{\prime}}\sin(\xi_j)+e^{-i\xi_j}].
\end{eqnarray}
\end{subequations}
\noindent The transmission coefficient for the first  sidebands, associated to the above exposition by  considering $-1\leq m-n \leq 1$   is in perfect agreement with numerical result, by considering  Floquet  sidebands  $m=0,\pm1 , ...,\pm12$, as displayed in Fig (\ref{shg}).
We must note that $t_0$ in the above  expressions denotes the transmission coefficient for the central band and the corresponding transmission coefficient  $T_0=|t_0|^{ 2}$ for the static barrier takes the  form:
\begin{equation}\label{10}
T_0=\frac{{b_{L,0}^{\prime}}^2(|F_0^{*}+W_0^{*}|^2)(|\gamma_0+\gamma_0^{*}|^2)}{(| W_0|^2 -| F_0|^2 )^2+4| F_0 W_0|^2  sin^2(q_0D)}.
\end{equation}
Consequently, in the static barrier and  for the cases  $(0,0)$ , $(1,- 1)$ and $(-1,1)$ at energy $E=0.5 V_0$,
the transmission is perfect  as well as independent of the incident angle and  barrier width.
\section{Numerical result and discussion}
As shown in the previous  section,  the transmission coefficient  for the first sidebands can be easily obtained analytically. On the other hand, a sufficiently large number of sidebands is needed to obtain the total transmission which can be calculated numerically. The total transmission as illustrated in Fig. (\ref{shga}-a:c) is independent of  $\alpha$, incidence angle and width of the barrier for the case $(0,0)$ and  $(1,-1)$   with $E=V_0/2$. However, in the case $(1,1)$, the change of $\alpha$ affects significantly the electron tunneling. Fig. (\ref{shga}-b,c) shows how the tunnelling is influenced by incidence angle and width of the barrier.  The resonances occur in the same  values of incidence angle and width of the barrier in all of following cases: $(\pm 1,0), (0,\pm 1), (1, 1)$.  However,  the resonant peaks  are characterized  by changing width of the barrier  for cases, $(\pm 1,0)$,  $(0,\pm 1)$    and $(1,1)$. For more details  we  show  the transmission coefficients $T_m$ for the  $(0,0)$ with $m=-3,...,3$ in Fig. (\ref{shga}-d) which is obtained by the numerical solution of the linear equation system  (11). We notice that for small value of $\alpha$ the transmission is mainly due to the central band.  In this  figure, we only show  several  sidebands, but in the  numerical results for the total transmission we used $m=-12,..,12$. 
 On the other hand, we have super Klein tunneling in static barrier and in the  cases $(0,0)$ , $(1,- 1)$ and $(-1,1)$, since the term $W_0 F_0 $ is equal to zero in Eq. (\ref{10}). 
 Independence of the the total transmission coefficient  in the presence of time-periodic potential  in the spacial cases  arises from this fact that absorbing or emitting $m$ photons have the same probability to cross the barrier.

\begin{figure}[t]
    \centering
    \subfigure[]{\includegraphics[width=0.4\textwidth]{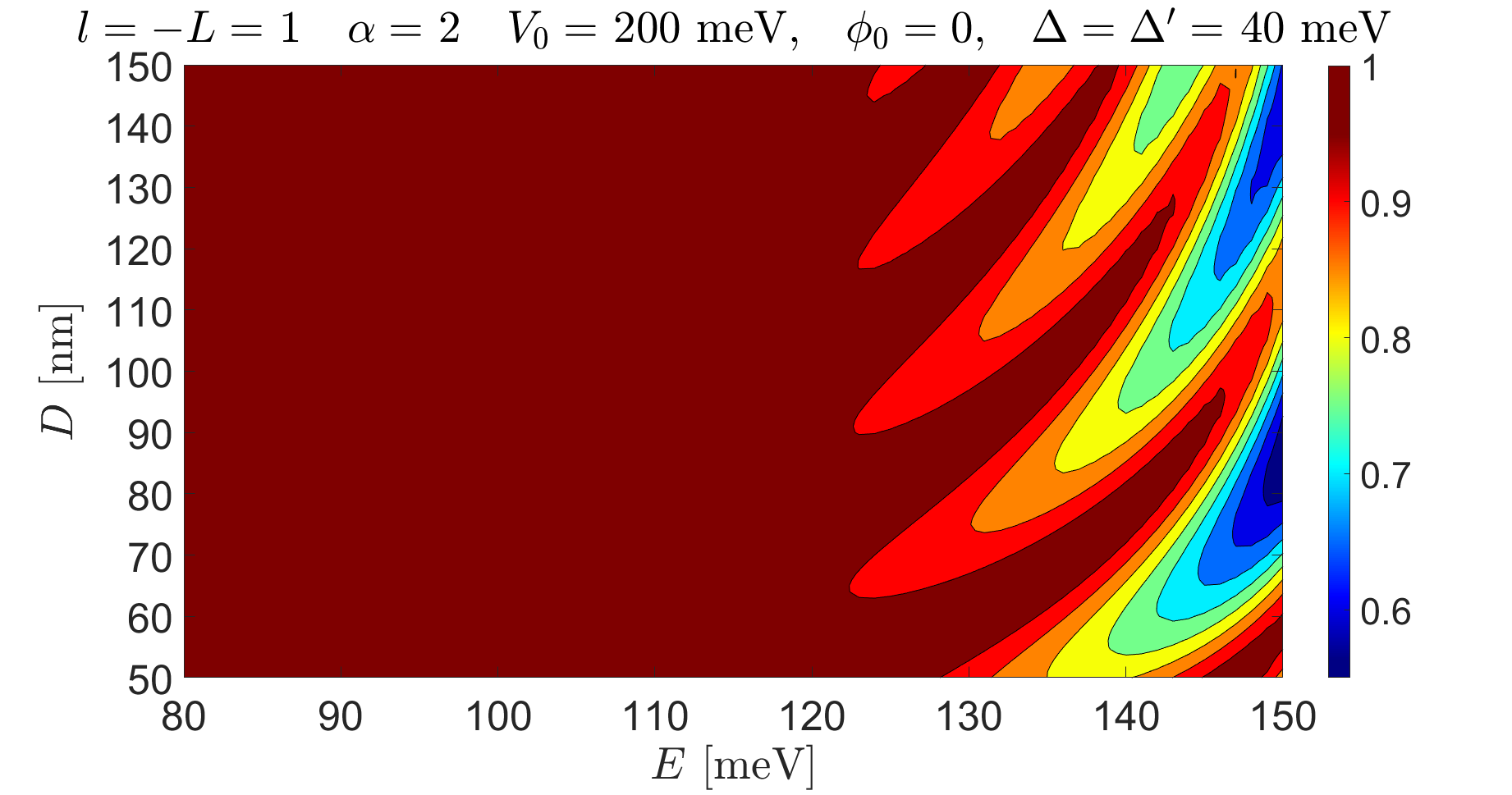}}\hspace{0.01em}%
    \subfigure[]{\includegraphics[width=0.4\textwidth]{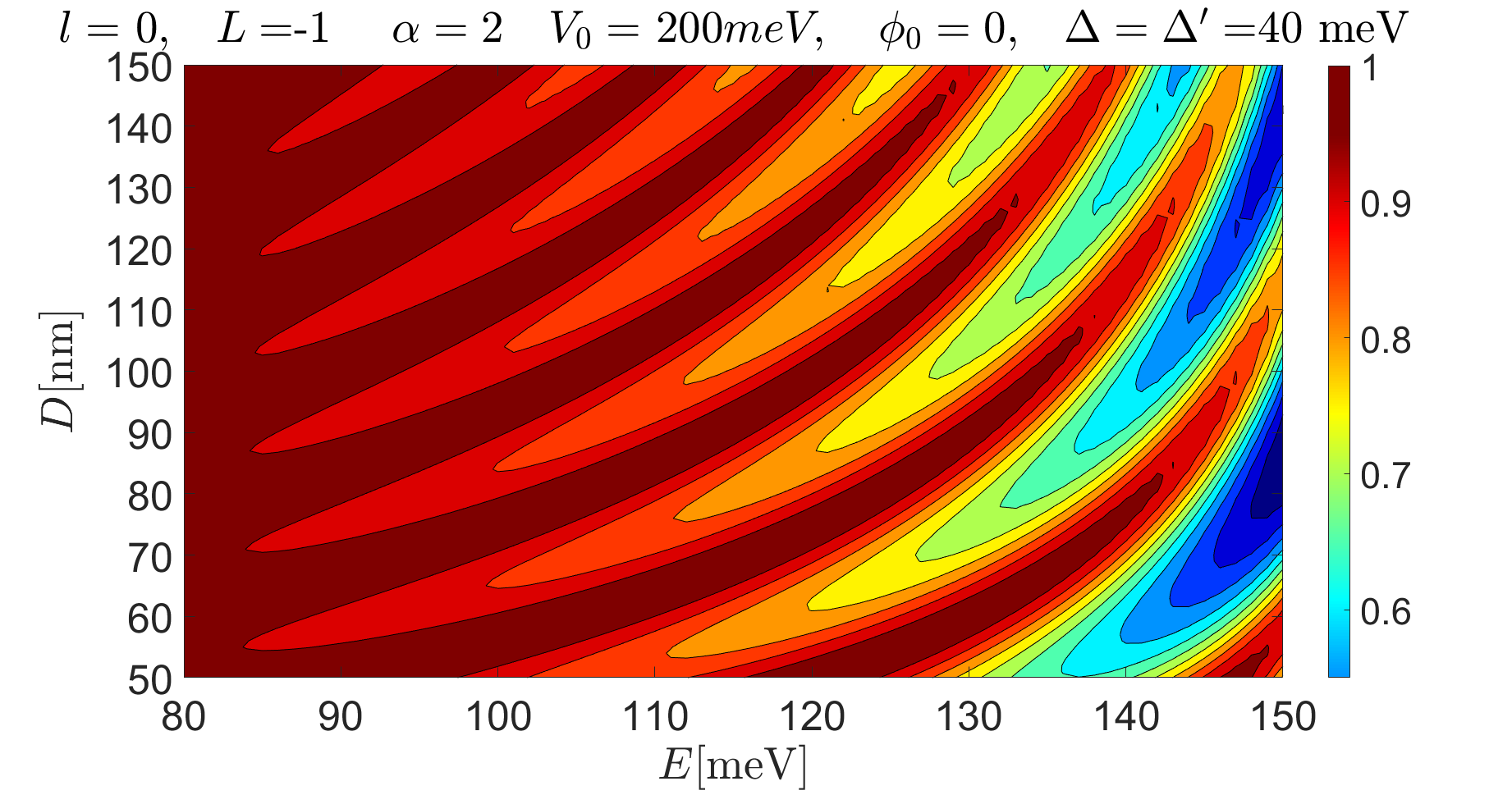}}
    \subfigure[]{\includegraphics[width=0.4\textwidth]{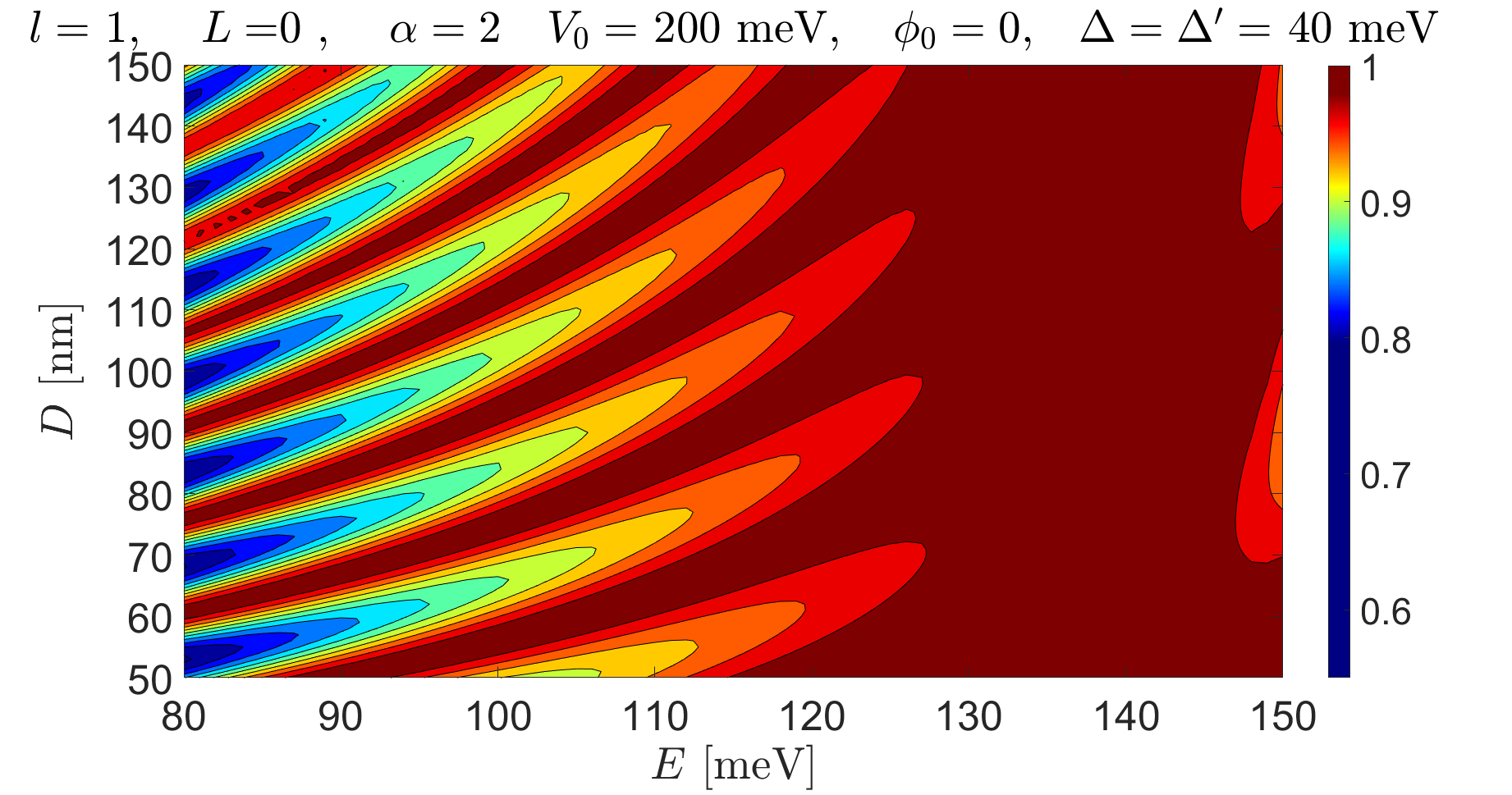}}\hspace{0.01em}%
    \subfigure[]{\includegraphics[width=0.4\textwidth]{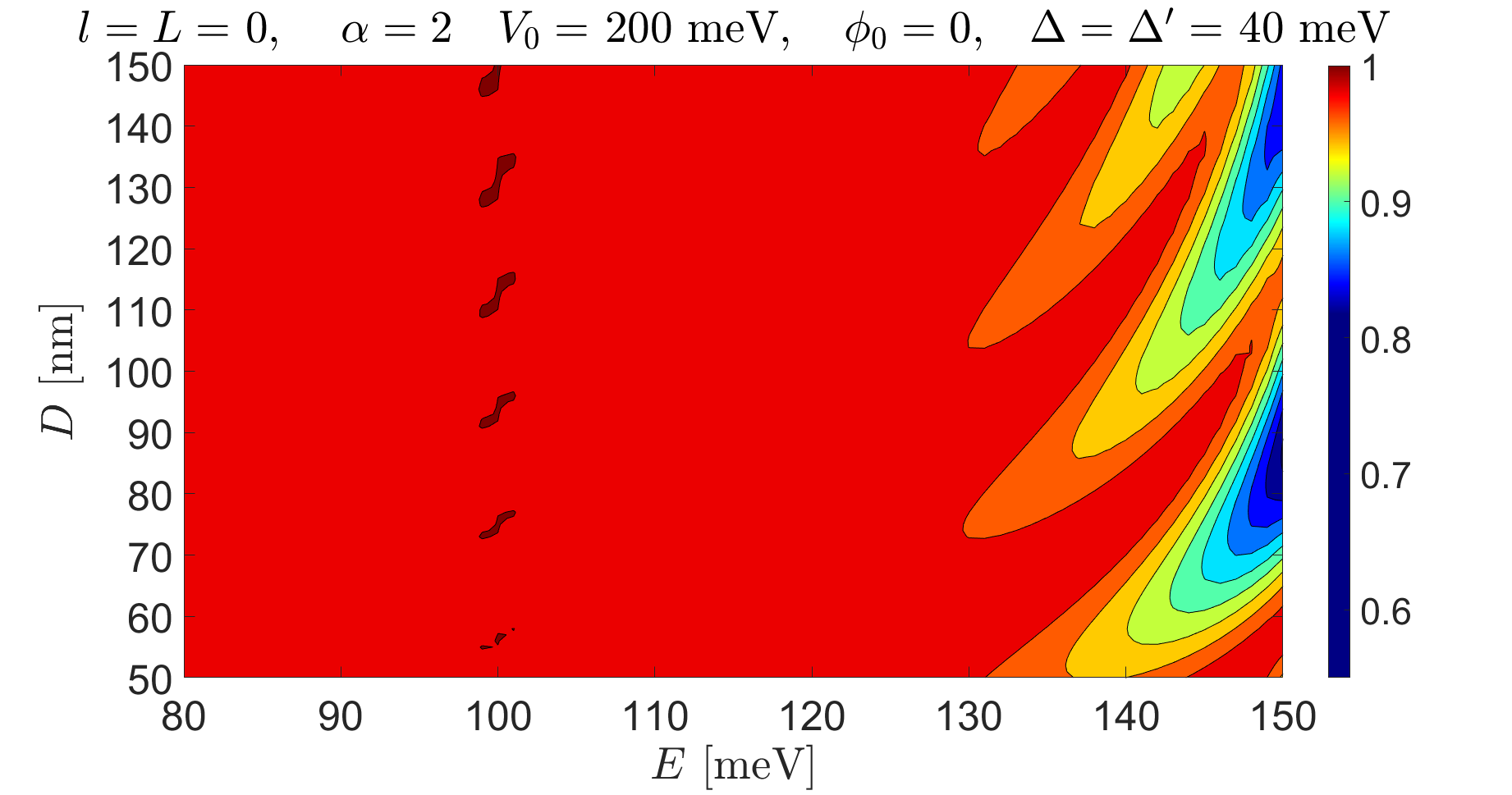}}
    \caption{  The total transmission  as functions of  barrier width  and energy with  $V_0=200$ meV,  $\alpha=2$ and $\Omega=1.25$ THz by considering $-12\leq m \leq 12$  with $\Delta=\Delta^ \prime$  for different cases:  (a) $(1,-1)$(b) $(0,-1)$ (c) $(1,0)$ (d) $(0,0)$. }
    \label{sh5}
\end{figure}
\noindent Figure  (\ref{sha3}) shows the angular dependence of the total transmission $T=\Sigma_{m=-u}^u T_m$ .  Moreover, one can easily see that  for  a specific value $V_0/E=2$ in cases $(1,-1)$ and $(0,0)$, applying time-periodic potential has no visible effect on super Klein tunneling for different incident angles.
We also notice that  increasing  energy ($E>0.5 V_0$) leads to  decreasing of the transmission  for large value of incident angle due to the increasing evanescent waves in the potential region. 
\begin{widetext}
\afterpage{\begin{figure}[t]
    \centering
    \subfigure[]{\includegraphics[width=0.3\textwidth]{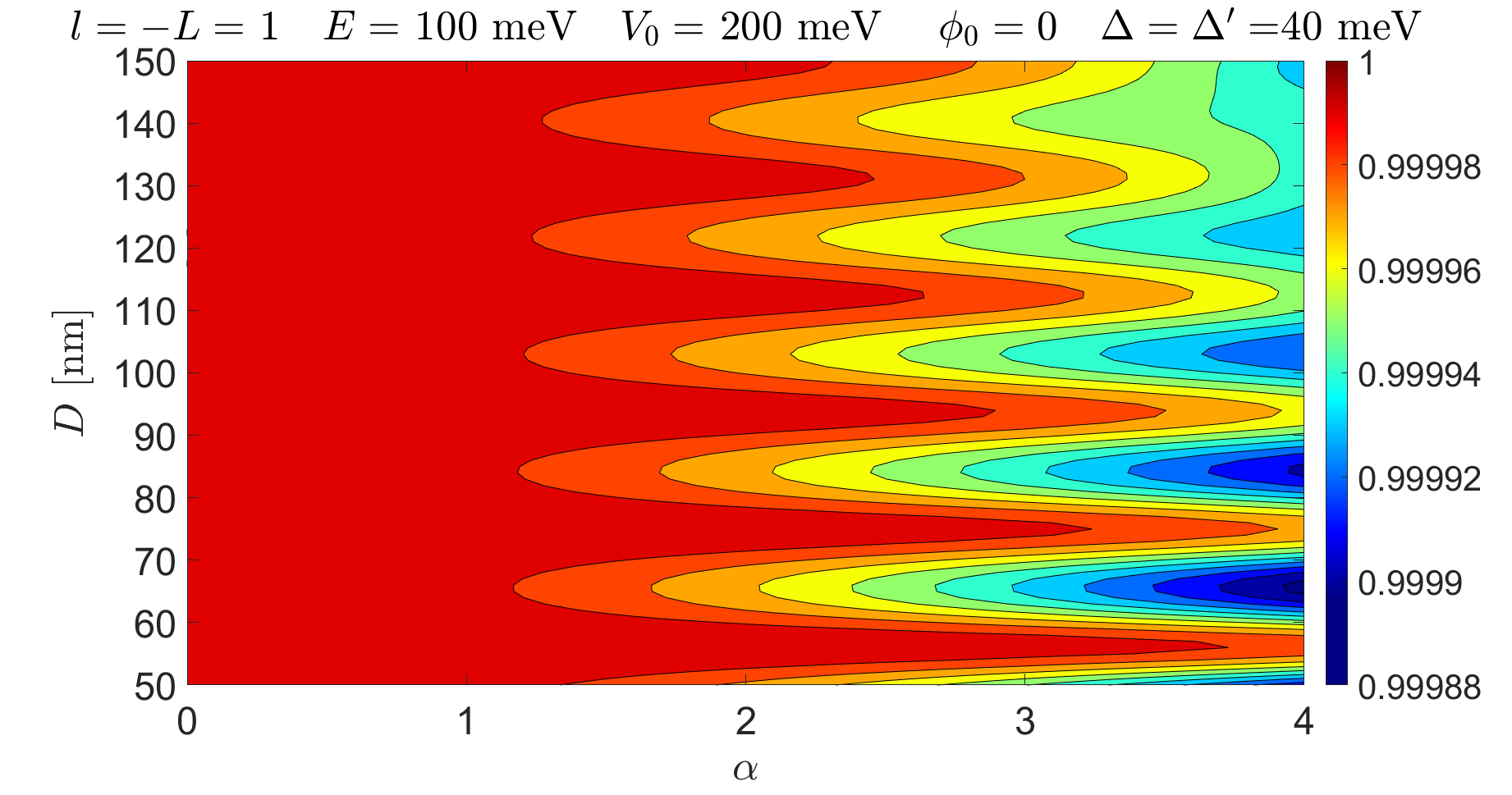}}\hspace{0.01em}%
    \subfigure[]{\includegraphics[width=0.3\textwidth]{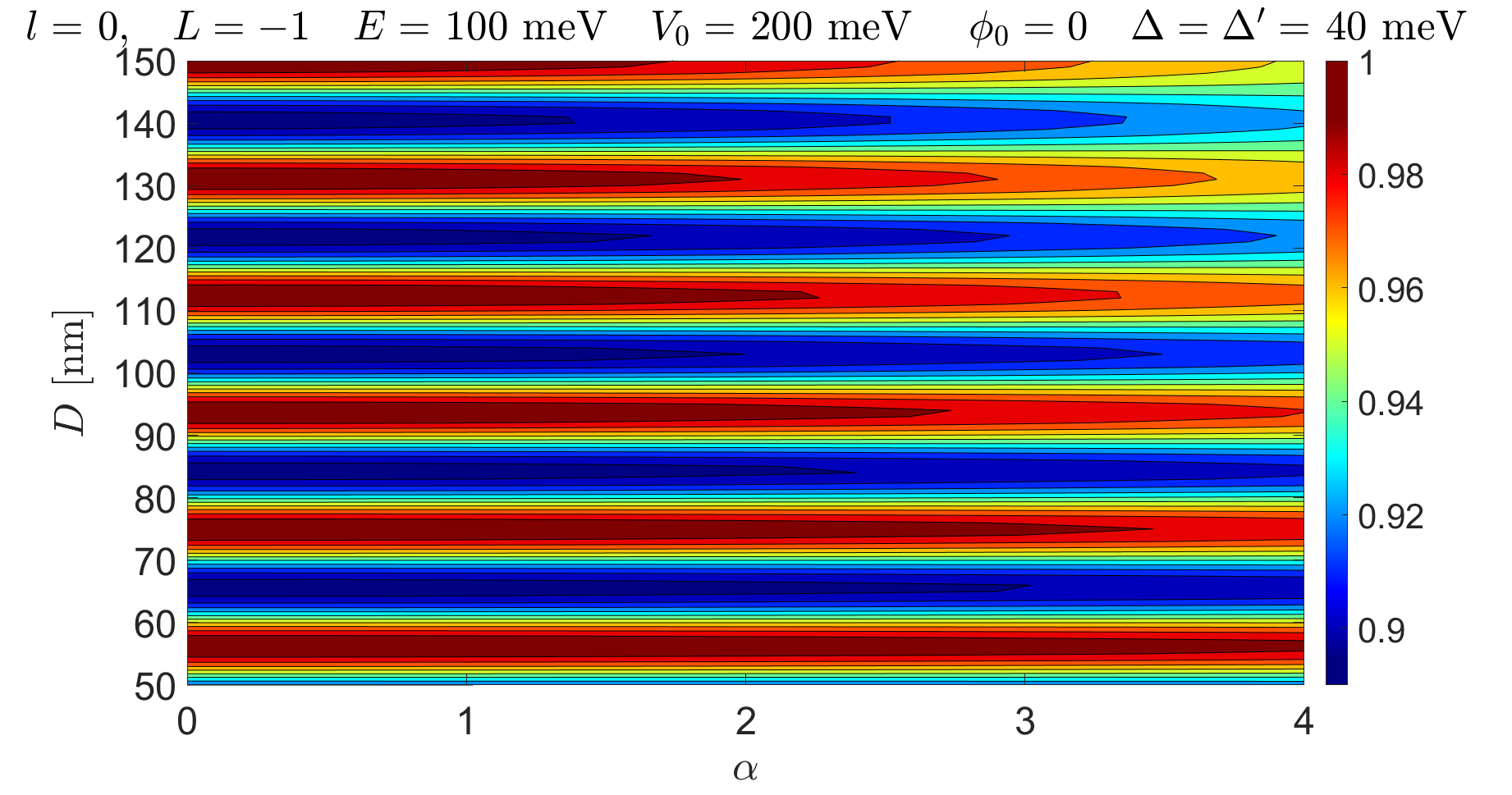}}
    \subfigure[]{\includegraphics[width=0.3\textwidth]{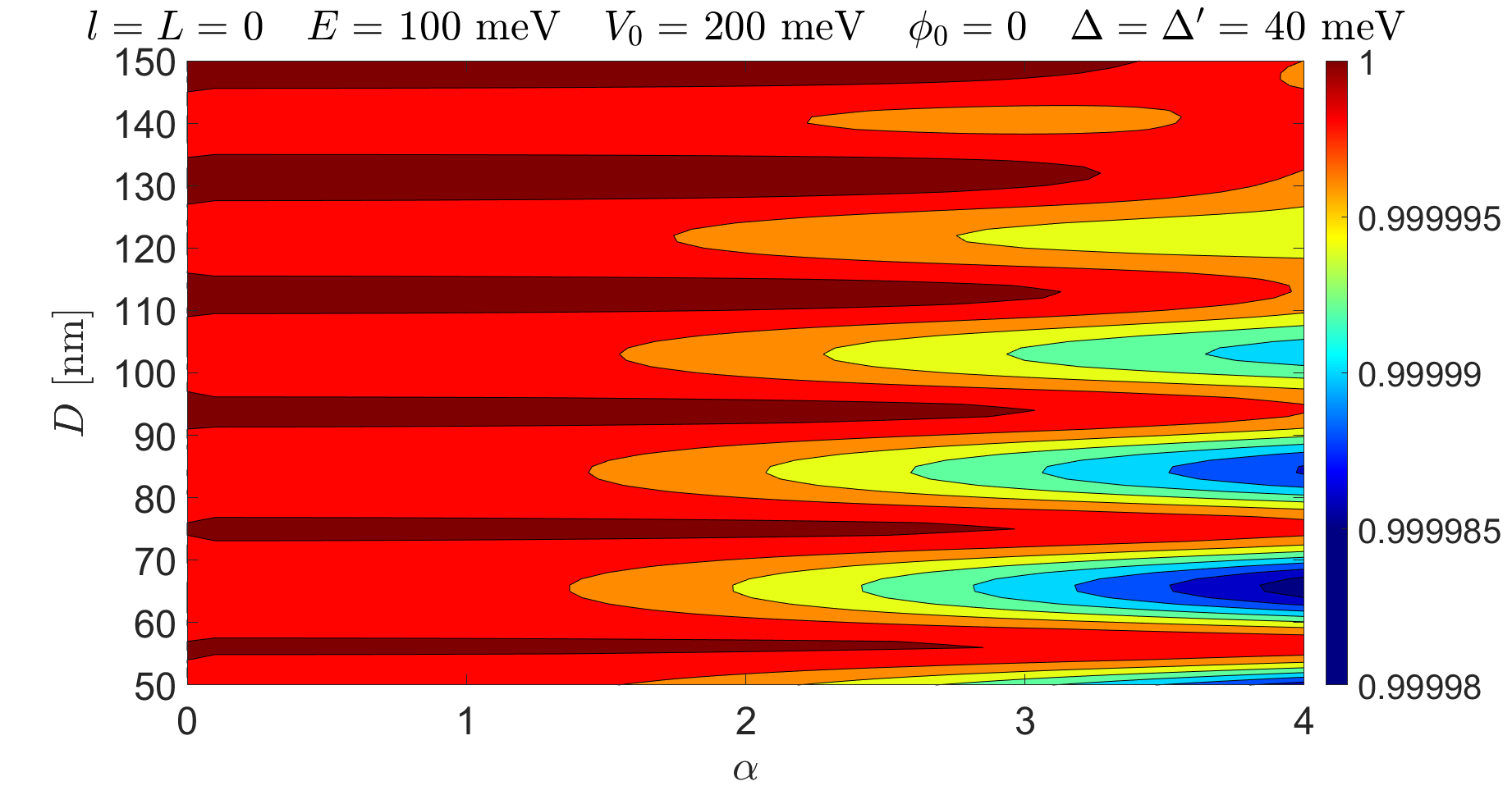}}
    \caption{  The total transmission  as functions of  width of barrier and  $\alpha$ with  $V_0=200$ meV, $\alpha=2$ and $\Omega = 1.25$ THz by considering  $-12 \leq m \leq 12 $ with $\Delta=\Delta^ \prime$ for different cases:  (a) $(1,-1)$(b) $(0,-1)$  (c) $(0,0)$  . }
    \label{sh6b}
\end{figure}
\begin{figure}[t]
    \centering
    \subfigure[]{\includegraphics[width=0.3\textwidth]{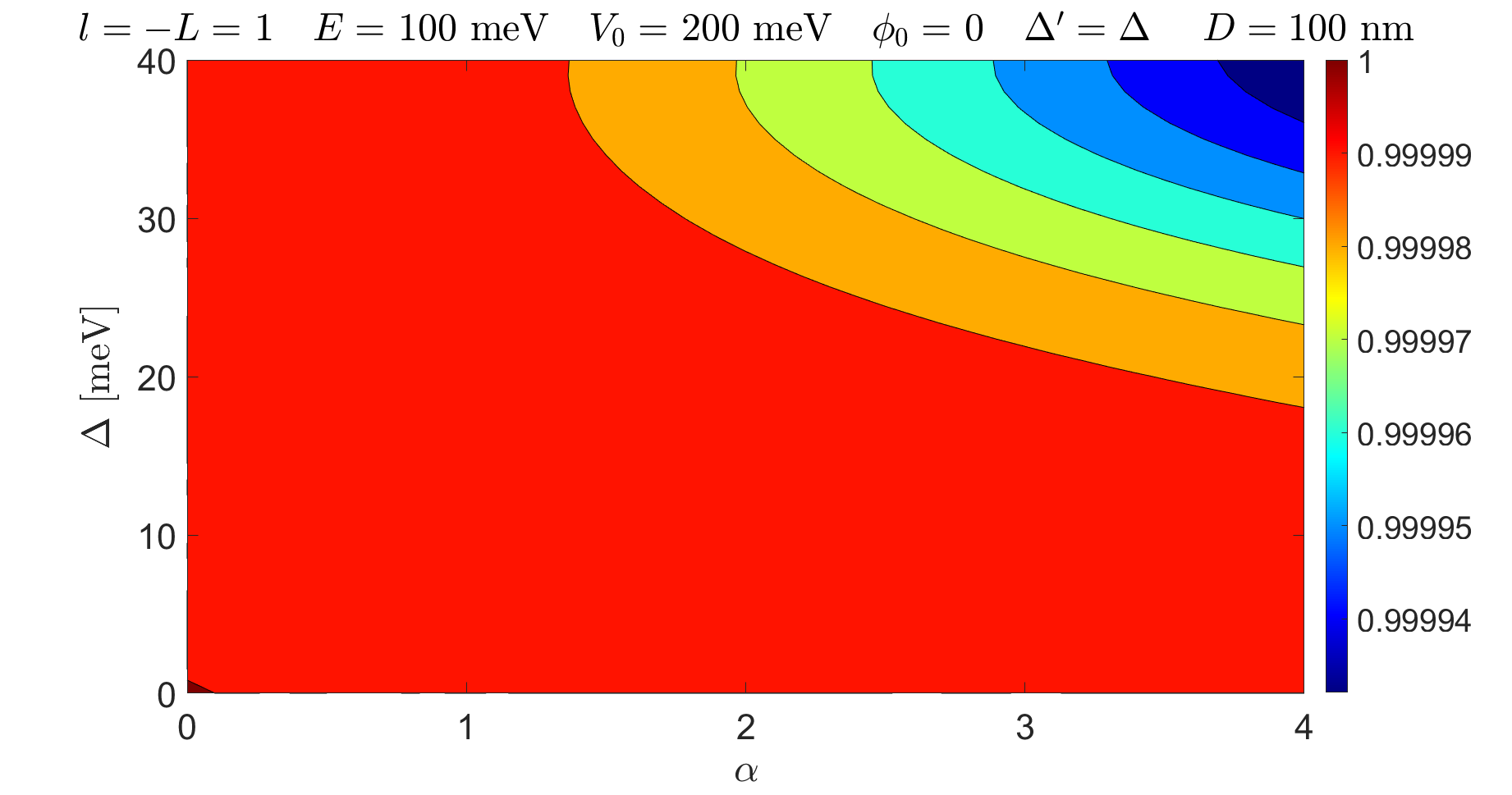}}
    \subfigure[]{\includegraphics[width=0.3\textwidth]{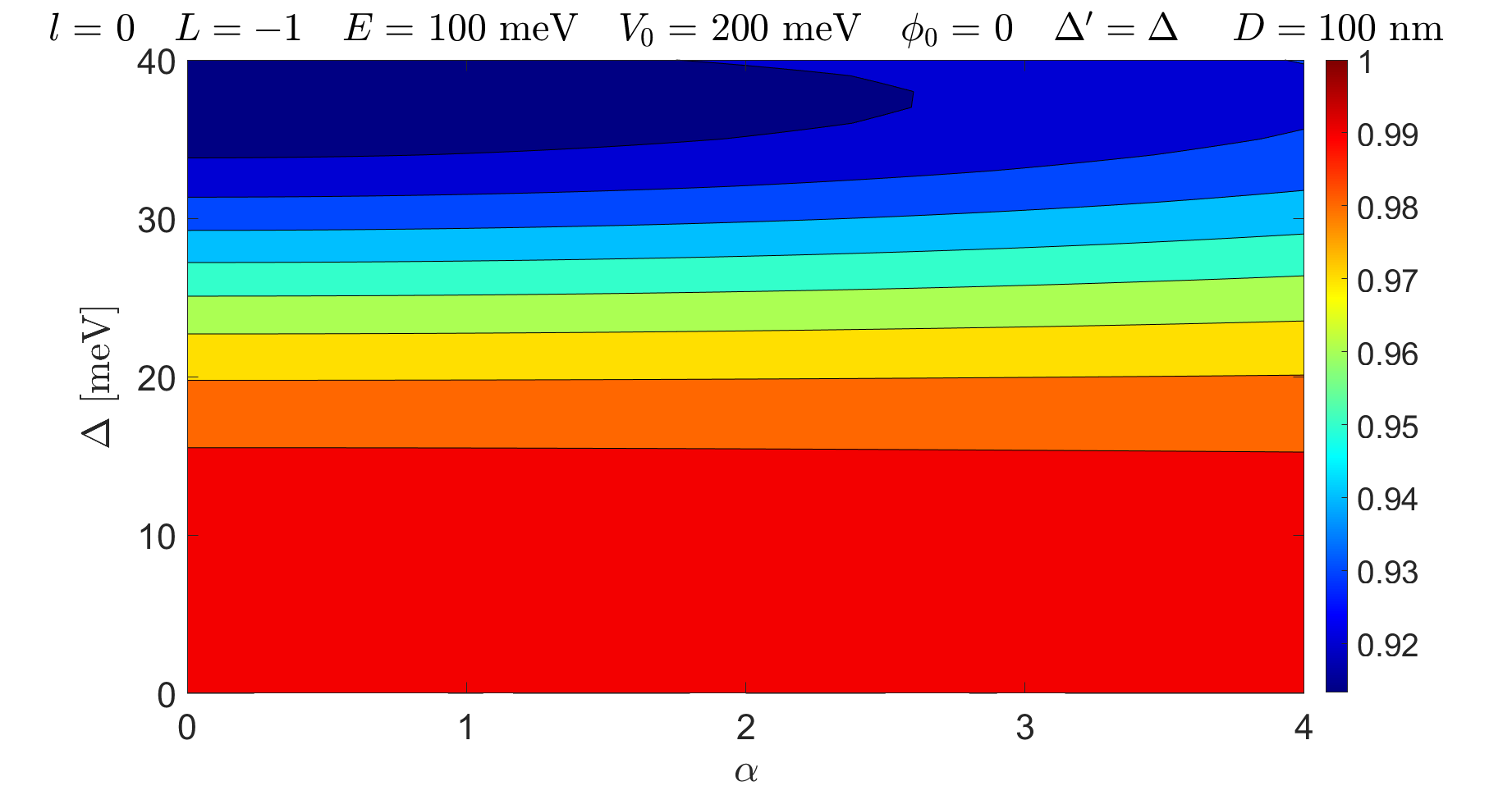}}
    \subfigure[]{\includegraphics[width=0.3\textwidth]{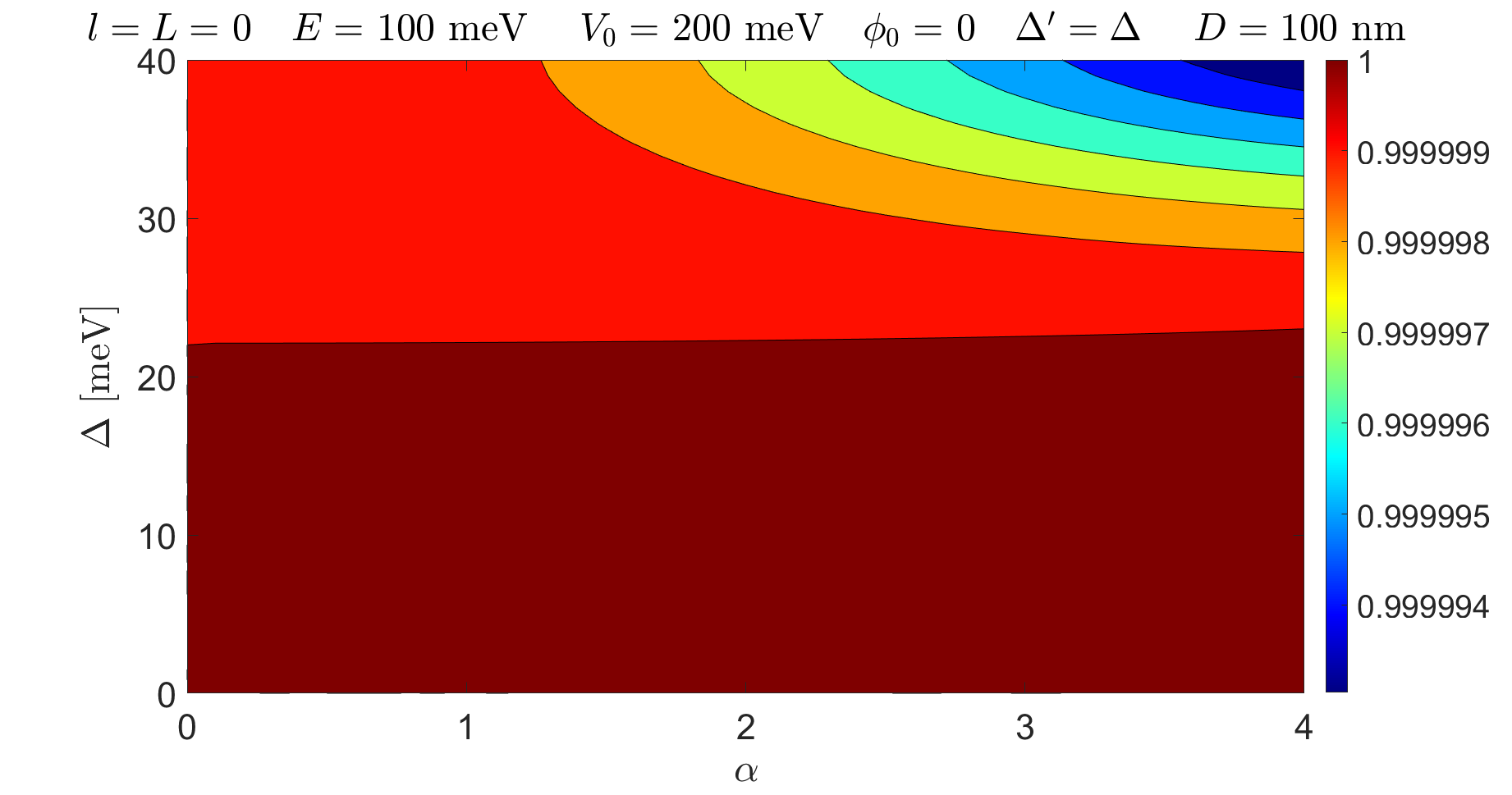}}
    \caption{  The total transmission  as functions of mass generation
term $\Delta$  and  $\alpha$ with  $V_0=200 meV$ ,$E=100 meV$, $\alpha=2$ and $\Omega = 1.25$ THz by considering $-12\leq m \leq 12$ with $\Delta=\Delta^ \prime$ for different cases:  (a) $(1,-1)$(b) $(0,-1)$  (c) $(0,0)$. }
    \label{sh7}
\end{figure}}

\end{widetext}
\noindent In  Fig. (\ref{sh5}), we obtain the total transmission probability for the normal incidence as functions of barrier width and energy. At first glance, one can recognize that by tuning the width of barrier and energy, resonances can be characterized. The important point to be highlighted is that  Klein tunneling for  normal incident electron in case $(1,-1)$,  where $V_0/E=2$, is independent of   barrier width.
\noindent The transmission coefficient for normal incident electron  is shown in Fig (\ref{sh6b}) as functions of width of barrier and $\alpha$. This figure highlights that tunneling  is almost independent of free parameters $D$, $\alpha$   for cases $(1,-1)$ and $(0,0)$ where $V_0/E=2$ and  $\Delta=\Delta^ \prime$ at normal incidence. Further, in the case  $(0,-1)$    we can see that resonances in the tunneling  occurs by tuning   $V_{ac}$ and width of the barrier. 
\noindent In order to quantify  the transmission properties, we explain the behavior of the transmission as a function of effective mass $m=\Delta/v_f^2$. As we can see, in Fig. (\ref{sh7})  the transmission coefficient
is again almost independent  in the cases $(1,-1)$ and $(0,0)$  at  energy $E=0.5V_0$ for normal incidence. It is clearly shown by these figures that the super Klein tunneling   in cases  $(1,-1)$, $(-1,1)$ and $(0,0)$ with  $\Delta=\Delta^ \prime$ and at  energy $E=0.5V_0$ is independent of the free parameters in our system. 
 On the other hand in the case $(0,-1)$,   increasing of the  gap opening, decrease the transmission probability.  The difference of results for different cases  arises from the  location of flat bands inside and outside the barrier. Consequently, location of flat bands plays  an important role in    the transmission probability.  For the case  $(1,-1)$ and $(-1 ,1)$ and with    $\Delta=\Delta^ \prime$ and $V_0/E=2$ due to the equality of $a_m=a'_{-m}$, $b_m=b'_{-m}$ and 
$c_m=c'_{-m}$,   the transmission coefficient for the  sideband $m$ is approximately equal to the $-m$  one. On the other hand,
 the transmission coefficient for side band $m$ is equal to $-m$ for the case $(0,0)$  with    $\Delta=\Delta^ \prime$ and $V_0/E=2$  due to the equality of $a_m=c'_{-m}$, $b_m=b'_{-m}$ and  $c_m=a'_{-m}$.   Consequently,  for the cases where absorbing or emitting m photons have the same probability to cross the barrier, the total transmission is independent of incidence angle, width of the barrier and the ratio $\alpha$.

\section{Conclusion and outlook}
In this paper we use  analytical and numerical techniques   to study  transmission probability for massive pseudo-spin one particles including  periodic time-dependent potential. Our analytical solutions are
generally in excellent agreement with numerical results. We explained the difference in the band gap opening by considering the cases corresponding to different locations of flat band.  We show that at energy $E=0.5V_0$ when the transmission coefficient for the sideband $m$ is approximately equal to the $-m$ one,  applying time periodic potential has no visible effect on super Klein tunneling, which is an important considerations for  the design of electronic devices. Furthermore, by tuning free parameters, we found  some particular values leading to transmission resonances.  The next result to be highlighted is that  increasing  energy ($E>0.5 V_0$) leads to  decreasing of the transmission  for large value of incident angle due to the increasing evanescent  modes. We find that a time-periodic barrier, beside the   standard  parameters such as  width and height of  the barrier, effective mass etc., provides a more flexible way for controlling the transmission probability. Helical wave-guides arranged in the dice lattice geometry can also be considered as a emulation of Floquet scattering of  pseudospin-one particles. Finally, our findings may have important impact on design and analysis of  Floquet topological insulators.

\section*{Acknowledgments}
I thank  Fran\c{c}ois Leyvraz, Thomas H. Seligman, Yonatan Betancur-Ocampo, Diego Espitia and Georg Jakob for helpful discussion and 
I  gratefully acknowledge a fellowship from UNAM-DGAPA, CONACYT Project Fronteras 952, UNAM–DGAPA PAPIIT IN113620 and CONACYT
Project 254515.

\bibliography{bibliography}
\end{document}